\documentclass[aps,prl,superscriptaddress,reprint,longbibliography,twocolumn]{revtex4-2}

\usepackage{amsfonts}
\usepackage{subfigure}
\usepackage{amsmath}
\usepackage{amssymb}
\usepackage{amsbsy} 
\usepackage{bbm}
\usepackage{epsfig}
\usepackage{graphicx}
\usepackage{epstopdf}
\usepackage{color, xcolor}
\usepackage{mathdots}
\usepackage{braket}
\usepackage{latexsym}
\usepackage{amsthm}
\usepackage{hyperref}

\renewcommand{\Re}{\operatorname{Re}}

\def\be{\begin{equation}} \def\ee{\end{equation}}
\def\bea{\begin{eqnarray}} \def\eea{\end{eqnarray}}

\usepackage{lipsum}

\begin{document}

\title{Fragile non-Bloch spectrum and unconventional Green's function}

\author{Fei Song}\thanks{These authors contributed equally to this work.}
\affiliation{Kavli Institute for Theoretical Sciences, Chinese Academy of Sciences, Beijing 100190, China}
\affiliation{Institute for Advanced Study, Tsinghua University, Beijing 100084, China}

\author{Hong-Yi Wang}\thanks{These authors contributed equally to this work.}
\affiliation{Department of Quantum Science and Engineering, Princeton University, Princeton, NJ 08544, USA}
\affiliation{Institute for Advanced Study, Tsinghua University, Beijing 100084, China}

\author{Zhong Wang}
\email{wangzhongemail@tsinghua.edu.cn}
\affiliation{Institute for Advanced Study, Tsinghua University, Beijing 100084, China}

\begin{abstract}



In non-Hermitian systems, it is a counterintuitive feature of the non-Hermitian skin effect (NHSE) that the energy spectrum and eigenstates can be totally different under open or periodic boundary conditions, suggesting that non-Hermitian spectra can be extremely sensitive to non-local perturbations. Here, we show that a wide range of non-Hermitian models with NHSE can even be highly sensitive to local perturbation under open boundary conditions. The spectrum of these models is so fragile that it can be significantly modified by adding only exponentially small perturbations on boundaries. Intriguingly, we show that such fragile spectra are quantified by the Green's function exhibiting unconventional V-shape asymptotic behaviors. Accordingly, bi-directional exponential amplification can be observed. As an interesting consequence, we find a real-to-complex transition of the bulk spectrum induced by exponentially small boundary perturbations. Finally, we reveal a hierarchy of the asymptotic behaviors of non-Hermitian Green's functions, which restricts the frequency range for the presence of unconventional Green's functions.
\end{abstract}
\maketitle
\emph{Introduction.--} Non-Hermitian physics naturally arises in  classical and quantum open systems \cite{ashida2020nonhermitian, bergholtz2021exceptional}. An intriguing phenomenon uncovered in non-Hermitian systems is the non-Hermitian skin effect (NHSE) \cite{yao2018edge,yao2018nonhermitian,kunst2018biorthogonal,lee2019anatomy,martinezalvarez2018nonhermitian, ghatak2020observation,xiao2020nonhermitian, helbig2020generalized,weidemann2020topological,wang2022nonhermitian,liang2022dynamic,zhang2022review,ding2022nonhermitian}. This effect means that most of the eigenstates of a non-Hermitian Hamiltonian with open boundary conditions are localized around the boundary. The finding of NHSE has spawned fruitful investigations ranging from open quantum systems to classical wave dynamics \cite{song2019nonhermitian,haga2021liouvillian,liu2020helical,yang2022liouvillian,xue2022nonhermitian,wang2021quantum,Hu2023burst,xiao2024burst,zhu2024burst,yi2020nonhermitian,longhi2020nonblochband,longhi2019nonbloch,longhi2019probing,xiao2021observation,hu2024geometric,mcdonald2020exponentiallyenhanced,wanjura2020topological,borgnia2020nonhermitian,xue2021simple,zhu2022anomalous,kawabata2023entanglement,shao2024nonHermitian}. In one dimension (1d), it is well known that the NHSE can be quantitatively described in the framework of non-Bloch band theory \cite{yao2018edge,yokomizo2019nonbloch,yokomizo2022Nonbloch,yang2020nonhermitian,kawabata2020nonbloch,deng2019nonbloch,hu2023non}. 

 

Compared to one dimension, the physics of NHSE is dramatically enriched in two and higher dimensions. Higher dimensions bring in ramified types of NHSE, such as geometry dependent NHSE \cite{zhang2022universal,zhang2023dynamical,fang2022geometrydependent,wang2022nonhermitiana}, higher-order NHSE \cite{lee2019hybrid,okugawa2020secondorder,kawabata2020higherorder,zou2021observation,zhang2021observation},  and the NHSE induced by lattice defects or  magnetic fields \cite{schindler2021dislocation,sun2021geometric,kawabata2021topological}. The non-Bloch band theory in 1d suggests that an efficient approach to characterize NHSE is to take complex-valued wave vector residing on the generalized Brillioun zone (GBZ) \cite{yao2018edge,yao2018nonhermitian,yokomizo2019nonbloch,yokomizo2022Nonbloch,yang2020nonhermitian}. However, finding the higher-dimensional GBZ is much more challenging than in 1d \cite{wang2024amoeba,hu2024topological,jiang2023dimensional,yokomizo2023nonbloch,kai2024edge,xiong2024non,zhang2024algebraic,xu2023two}.


A central question of non-Hermitian band theory is whether, and in what sense, the GBZ concept can be unambiguously defined in higher dimension irrespective of the geometry and boundary details. To this end, we study representative models whose spectra are dramatically affected by locally perturbing the systems.  Their spectra and GBZ are fragile in the sense that they can be drastically modified by adding exponentially small local perturbations on the boundary. This appears to deviate from the prediction of the amoeba formulation of non-Bloch band theory because the latter is insensitive to such perturbations \cite{wang2024amoeba}. We find that this fragility is characterized by the unconventional asymptotic behaviors of the Green's function $G(\omega)=(\omega-H)^{-1}$ \footnote{Without extra explanations, the $\omega$ in $\omega-H$ denotes the product between $\omega$ and the identity matrix.}, where $H$ is the real-space Hamiltonian of the system. Furthermore, we find that the amoebic spectrum defined in Ref. \cite{wang2024amoeba} gives the maximal range of the frequency $\omega$ with such spectral fragility. This explains why the amoebic spectrum is the stable spectrum in the presence of generic local perturbations \cite{wang2024amoeba}.


\begin{figure}[ht]
	\includegraphics[width=8cm]{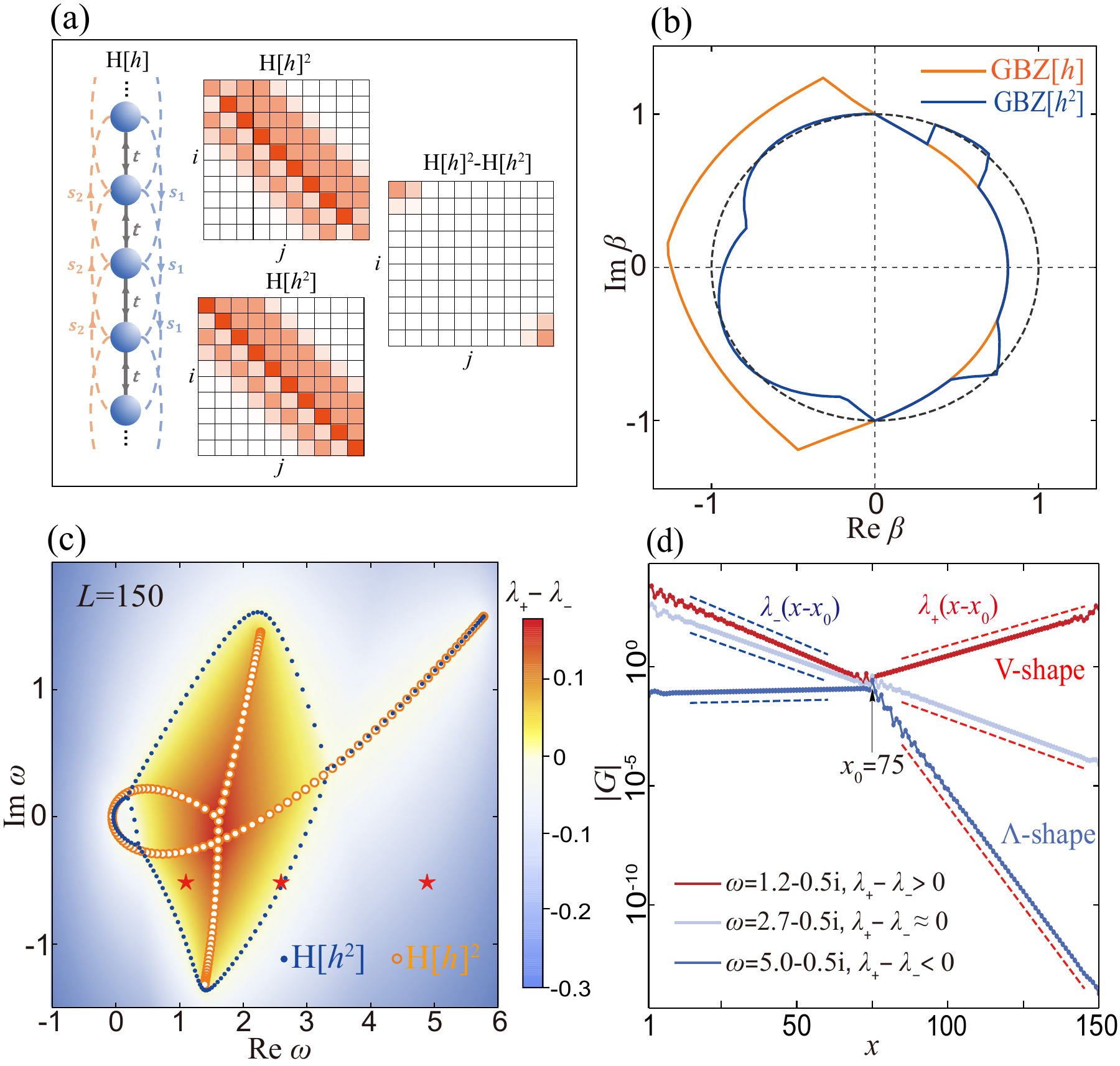}
\caption{The left side of (a) shows the real-space hopping of the model Eq.~(\ref{h}). The right side of (a) illustrates the matrix elements of the Hamiltonians $H[h]^2$ and $H[h^2]$, and their difference, where the opacity reflects the magnitude of the matrix elements. (b) shows the two GBZs determined by $h(k)$ and $h^2(k)$. The orange and blue dots in (c) compose the spectrum of Hamiltonians $H[h]^2$ and $H[h^2]$, respectively. The color map in (c) shows $\lambda_+-\lambda_-$, where coefficients $\lambda_+$ and $\lambda_-$ are obtained via fitting $\ln |G_{H[h]^2}(x,x_0=75; \omega))|$ with  $x\in[85,135]$ and $x\in[15,65]$, respectively. Three types of typical Green's functions $G_{H[h]^2}(x,x_0; \omega))$ are shown in (d). The parameters are $t=1,s_1=0.5,s_2=0.2i$.} \label{figure_1d}
\end{figure}


\emph{A 1d example.--} Our work reveals that in various non-Hermitian models with open boundary conditions, even if the real-space Hamiltonian remains unchanged in the bulk, the spectrum can be sharply changed in response to the local alterations on the boundary. As a typical example, we  start from comparing the spectra of two 1d Hamiltonians $H[h]^2$ and $H[h^2]$. Here, $H[h]$ denotes the real-space Hamiltonian corresponding to the Bloch Hamiltonian $h(k)$, whose elements are given by $(H[h])_{ij}=\int_0^{2\pi} (dk/2\pi) h(k) e^{ik(i-j)}$, and $H[h^2]$ follows a similar expression. For concreteness, we take \bea
h(k)=2t \cos k+s_1 e^{i2k}+s_2 e^{-i2k}. \label{h}
\eea
As illustrated by Fig.~\ref{figure_1d}(a), the difference $B=H[h]^2-H[h^2]$ has non-zero elements $B_{ij}$ only when both indices $i,j$ are close to the left or right end of a finite 1d chain. Since $H[h]^2$ and $H[h^2]$ differ from each other only at system's boundaries, one might expect that they should have the same bulk spectrum.


However, the actual diagonalization shows that the spectra of $H[h]^2$ and $H[h^2]$ are very different [Fig.~\ref{figure_1d}(c)]. Theoretically, these two spectra can be  predicted by the analytic extension $h^2(k\rightarrow -i\ln \beta)$, and taking $\beta$ on different GBZs. For the spectra of $H[h]^2$ and  $H[h^2]$, the GBZ is determined by the Bloch Hamiltonian $h(k)$ and $h^2(k)$, respectively. According to the non-Bloch band theory in 1d, GBZ is solved by requiring the middle two roots $\beta_i(E)$ of the characteristic equation $\det [E-\tilde{h}(k\rightarrow -i\ln \beta]=0$ to have the same moduli \cite{yao2018edge,yokomizo2019nonbloch,yokomizo2022Nonbloch,yang2020nonhermitian}. Here, $\tilde{h}(k)$ should be taken as $h(k)$ and $h^2(k)$. From Figs.~\ref{figure_1d}(b)(c), we find a partial overlap between the two GBZs and their corresponding spectra. Then, what causes the difference between the overlapping and non-overlapping parts?

We find that this difference is related to the asymptotic behaviors of the Green's function $G_{H[h]^2}(x,x_0; \omega)=\langle x|(\omega-H[h]^2)^{-1}|x_0\rangle$. The modulus of the Green's function $|G_{H[h]^2}(x,x_0; \omega)|$ exponentially grows or decays as $e^{\lambda(x-x_0)}$ when the distance $|x-x_0|$ is sufficiently large. We denote the coefficient as $\lambda_{+}$ for $x>x_0$ and as $\lambda_-$ for $x<x_0$. As shown in Fig.~\ref{figure_1d}(c), the Green's function $G_{H[h]^2}(x,x_0; \omega)$ exhibits $\lambda_+-\lambda_->0$ in the frequency region colored in orange while exhibits $\lambda_+-\lambda_-<0$ in the frequency region colored in blue. Moreover, we demonstrate the behaviors of the Green's function with representative frequencies in Fig.~\ref{figure_1d}(d). We see that the dependence of $\ln |G_{H[h]^2}(x,x_0; \omega)|$ on the position $x$ is V-shaped when $\lambda_+-\lambda_->0$, and $\Lambda$-shaped when $\lambda_+-\lambda_-<0$. 


The $V$-shape asymptotic behavior is \emph{ad hoc}, since it would imply signals being amplified whichever direction they travel. Even in 1d non-Bloch band theory, $V$-shape is not allowed because $\lambda_\pm$ are given by the middle two roots of the characteristic equation \cite{xue2021simple}. For example, $\lambda_+=\ln |\beta_n(\omega)|$ and $\lambda_-=\ln |\beta_{n+1}(\omega)|$ for the single-band model with the Bloch Hamiltonian $h(k)=\sum_{s=-n}^m t_s e^{iks}$, where $\beta_{n,n+1}(\omega)$ are the roots of $\omega=h(k\rightarrow -i\ln \beta)$ sorted as $|\beta_1(\omega)|\leq\ldots\leq|\beta_{m+n}(\omega)|$. Obviously, this prediction always guarantees $\lambda_{+}-\lambda_{-}\leq0$ and thus forbids the presence of the $V$-shape behavior. Here, the V-shape behaviors of the Green's function $G_{H[h]^2}(x,x_0; \omega)$ stem from the special structure of the Hamiltonian $H[h]^2$, which allows the factorization
\bea
\frac{1}{\omega-H[h]^2}=\frac{1}{2\sqrt{\omega}}\left(\frac{1}{\sqrt{\omega}-H[h]}+\frac{1}{\sqrt{\omega}+H[h]}\right).
\eea
Notice that the asymptotic behaviors of Green's function $G_{H[h]}(\omega)=(\omega-H[h])^{-1}$ still obeys the usual prediction. Then, from the above factorization, the coefficients $\lambda_\pm$ for $G_{H[h]^2}(x,x_0; \omega)$ can be predicted as $\lambda_+=\max\{\beta_2(\sqrt{\omega}),\beta_2(-\sqrt{\omega})\}$ and $\lambda_+=\min\{\beta_3(\sqrt{\omega}),\beta_3(-\sqrt{\omega})$, where $\beta_{2,3}(\pm\sqrt{\omega})$ are middle two roots of the equations $\pm \sqrt{\omega}=h(k\rightarrow -i\ln \beta)$. Consequently, V-shape behaviors arise when the condition like $\beta_2(-\sqrt{\omega})\leq \beta_3(-\sqrt{\omega})< \beta_2(\sqrt{\omega})\leq \beta_3(\sqrt{\omega})$ is fulfilled. 


\begin{figure*}
	\includegraphics[width=16cm]{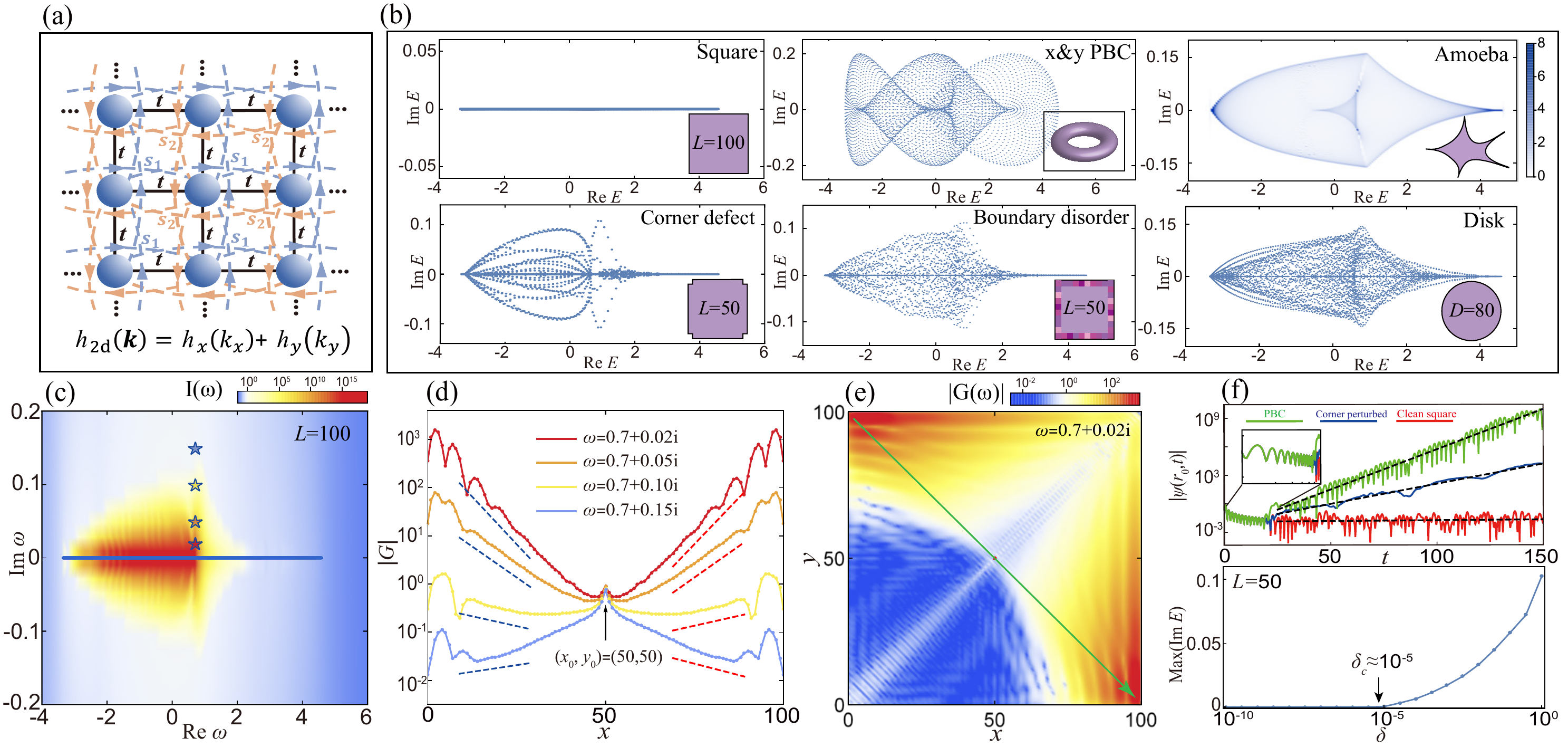}
\caption{(a) The real-space hopping of the model Eq.~(\ref{Bloch_2d}) with $h_x(k)=h_y(k)=2t\cos k+s_1e^{ik}+s_2e^{-ik}$. 
(b) Subfigures show the spectra calculated on clean square $H_{2d}$, PBC, the amoeba formulation, square without the four corners, square with boundary disorder ($H=H_{2d}+\sum_{\mathbf{r}\in \text{boundary}}V(\mathbf{r})|\mathbf{r}\rangle\langle\mathbf{r}|$ with random $V(\mathbf{r})\in[-1/2,1/2]$), and disk, respectively. 
(c) Color map of the V-shape proxy $I(\omega)$ obtained on a clean square. (d) Typical behaviors of the Green's function $\langle x,y|G_{2d}(\omega)|x_0=50,y_0=50\rangle$, where $(x,y)$ is taken along the green arrow depicted in (e). (e) Color map for the Green's function $|\langle x,y|G_{2d}(\omega)|x_0=50,y_0=50\rangle|$ with $\omega=0.7+0.02i$. (f) Upper panel: the evolution of $\langle x_0,y_0|\psi(t)\rangle$. The red and blue lines are obtained with $H=H_{2d}$ and $H=H_{2d}+\sum_{\mathbf{r}\in \text{four corners}}|\mathbf{r}\rangle\langle\mathbf{r}|$, respectively, and the green line with PBC. For all three cases, $L=50$ and the initial state is $|\psi(0)\rangle=|x_0=25,y_0=25\rangle$. Lower panel: the maximal imaginary part of the spectrum of $H_{2d}+\delta\sum_{\mathbf{r}\in \delta\text{four corners}}|\mathbf{r}\rangle\langle\mathbf{r}|$. One can find a real-to-complex transition that happens at $\delta_c\approx 10^{-5}$. The parameters are $t=1,s_1=0.2,s_2=0.1$. }\label{figure_2d}
\end{figure*}

\emph{V-shape Green's functions and spectral instability.--} Let us understand why V-shape Green's functions are vital for the spectral difference between $H[h]^2$ and $H[h^2]$. We recall that in analogy to the electron gas in 2d, the spectral density $\rho(\omega)=(1/L)\sum_{n=1}^L\delta^2(\omega-E_n)$ of the eigenenergies of the Hamiltonian $H$ with the dimension $L$ can be derived from the Laplacian of a Coulomb potential given by 
\bea
\Phi(\omega)=\frac{1}{L}\sum_{n=1}^L\log |\omega-E_n| =\frac{1}{L}\Re \text{Tr} \log \left(\omega-H\right). \label{potential}
\eea
The Green's function  $G(\omega)=(\omega-H)^{-1}$ appears in the change of this potential as $\delta \Phi(\omega)=(1/L)\Re\sum_{k=1}^{+\infty}\text{Tr} [G(\omega)\delta B]^k/k$ when the Hamiltonian is perturbed from $H$ to $H-\delta B$. Therefore, analyzing $\delta \Phi(\omega)$ can reveal a connection between the behaviors of the Green's function and the spectral stability under the perturbation $\delta B$.

To understand the difference between the spectra of $H[h]^2$ and $H[h^2]$, we study $H[h]^2$ with perturbation of the form $\delta B=\delta (|1\rangle\langle 1|+|L\rangle\langle L|)$, which is simple while capturing the locality of $H[h]^2-H[h^2]$ (see Fig.~\ref{figure_1d}(c)). When $H[h]^2$ is perturbed by $\delta B$, the first-order correction of the potential is $\delta \Phi^{(1)}(\omega)=(\delta/L)\Re [G_{H[h]^2}(1,1;\omega)+G_{H[h]^2}(L,L;\omega)]$, which is always $\mathcal{O}(1/L)$ due to the bi-orthogonality of eigenstates \footnote{According to Ref. \cite{kunst2018biorthogonal}, the bi-orthogonality between right eigenstates $|nR\rangle$ and left eigenstates $|nL\rangle$ of a non-Hermitian Hamiltonian $H$ makes sure that $\langle x|nR\rangle\langle nL|x\rangle$ is extended for the bulk states. One can use this property to show that the diagonal elements of the Green's function $G(\omega)=(\omega-H)^{-1}$ are at order $\mathcal{O}(1)$.}. In contrast, higher-order corrections are exponentially large when the Green's function $G_{H[h]^2}(\omega)$ shows V-shape asymptotic behaviors. For example, the second-order correction $\delta \Phi^{(2)}(\omega)=(\delta^2/2L)\Re [G_{H[h]^2}(1,1;\omega)^2+G_{H[h]^2}(L,L;\omega)^2+2G_{H[h]^2}(1,L;\omega)G_{H[h]^2}(L,1;\omega)]\sim \delta^2\left[1+e^{(\lambda_+-\lambda_-)L}\right]/L$, which is exponentially large for $\lambda_+-\lambda_->0$ (V-shape), while  is at order $\mathcal{O}(1/L)$ for $\lambda_+-\lambda_-<0$ ($\Lambda$-shape). As a result, $\delta \Phi(\omega)$ with V-shape Green's functions fails to converge for an exponentially small $\delta$, whereas the $\delta \Phi(\omega)$ with $\Lambda$-shape Green's functions vanishes in the thermodynamic limit ($L\rightarrow+\infty$). This implies that whether a spectrum is unstable or stable under local perturbations near the boundary depends on whether the Green's function $G(\omega)$ is V-shape or $\Lambda$-shape for the frequency $\omega$. This finding explains the results in Fig.~\ref{figure_1d}(c).  

The connection between V-shape Green's functions and spectral instability is general and holds for multi-band and higher-dimensional systems. A multi-band analog of the phenomenon discussed here can be found in the model with the critical NHSE \cite{li2020critical,SM}. Subsequently, we will continue to verify this connection in 2d.


\emph{V-shape Green's functions in 2d.--} We show the manifestation of V-shape Green's functions in a category of 2d models with the Bloch Hamiltonian
\bea
h_{2d}(k_x,k_y)=h_x(k_x)+h_y(k_y). \label{Bloch_2d}
\eea
As illustrated in Fig.~\ref{figure_2d}(a), this Bloch Hamiltonian describes the model with independent real-space hoppings along $x$ and $y$ directions.  A special property of such a model is that when it is placed on a rectangular geometry with the basis $\{|x\rangle\otimes|y\rangle,x\in[1,L_x],y\in[1,L_y]\}$, its real-space Hamiltonian becomes the Kronecker sum of two 1d Hamiltonians with open boundary conditions, which can be written as
\bea
H_{2d}=H[h_x]\otimes I_{L_y}+I_{L_x}\otimes H[h_y].\label{real_2d}
\eea
This property allows  $H_{2d}|\psi_{mn}\rangle=E_{mn}|\psi_{mn}\rangle$ to be solved by $E_{mn}=E_{m,x}+E_{n,y}$ and $|\psi_{mn}\rangle=|\phi_m\rangle \otimes|\xi_n\rangle$, where $E_{x,m},E_{y,n}$ and $|\phi_m\rangle, |\xi_n\rangle$ satisfy $H[h_x]|\phi_m\rangle=E_{x,m}|\phi_m\rangle$ and $H[h_y]|\xi_n\rangle=E_{y,n}|\xi_n\rangle$. A similar result exists for the left eigenstates satisfying $H_{2d}|\tilde{\psi}_{mn}\rangle=E^*_{mn}|\tilde{\psi}_{mn}\rangle$. Then, akin to the 1d cases, the spectrum of $H_{2d}$ in the thermodynamic limit can be predicted by substituting the Bloch wavevector $(e^{ik_x},e^{ik_y})$ in Eq.~(\ref{Bloch_2d}) to the complex wavevector $(\beta_x\in \text{GBZ}_x,\beta_y\in \text{GBZ}_y)$, where $\text{GBZ}_{x(y)}$ denotes the GBZ of the 1d Hamiltonian $H[h_{x(y)}]$. This also means that the 2d GBZ of $H_{2d}$ obeys the structure $\text{GBZ}_{2d}=\text{GBZ}_x\otimes\text{GBZ}_y$.

However, this direct solution of 2d GBZ generally violates the prediction of the amoeba formulation of non-Bloch band theory. The spectrum given by $\text{GBZ}_{2d}=\text{GBZ}_x\otimes\text{GBZ}_y$ can be purely real, when $h_x(k_x)$ and $h_y(k_y)$ take the form of Eq.~(\ref{h}) with real parameters and their corresponding GBZs are smooth \cite{hu2024geometric}. In contrast, according to the amoeba formulation, a complex spectrum is obtained for the same Bloch Hamiltonian. The difference between these two spectra is notable in Fig.~\ref{figure_2d}(b). In  addition, this figure also shows the spectra calculated with other boundary conditions and geometries. We find that the spectrum under the square geometry is so fragile that even cutting off the four corners of the square can break its realness. Moreover, the spectrum tends to approach the amoebic spectrum when the square geometry is messed up by the disorder on its boundary or when the system is placed on a disk [Fig.~\ref{figure_2d}(b)].


Similar to the aforementioned 1d example, the fragile spectrum  of $H_{2d}$ is also accompanied by the V-shape behaviors of the Green’s function $G_{2d}(\omega)=(\omega-H_{2d})^{-1}$. We define a quantitative proxy for the V-shapedness in 2d as
\bea
I(\omega)=\max_{\mathbf{r},\mathbf{r}'\in\text{boundary}}|\langle \mathbf{r}'|G_{2d}(\omega)|\mathbf{r}\rangle\langle \mathbf{r}|G_{2d}(\omega)|\mathbf{r}'\rangle|.
\eea
Aligned with the definition in 1d, we call the Green's function V-shape when this indicator is exponentially large as $I(\omega)\sim \mathcal{O}(e^{L})$, where $L$ denotes the typical length scale of the 2d system with $\mathcal{O}(L^2)$ sites. As shown in Figs.~\ref{figure_2d}(c)(d), the Green’s function $G_{2d}(\omega)$ exhibits V-shape behaviors at the frequency $\omega$ that is close to the fragile spectrum. This can be understood as follows. When $\omega$ is sufficiently close to the real energy $E$, the Green's function $G_{2d}(\omega)$ and the projector $P(E)=\sum_{m,n, E_{mn}=E} |\psi_{mn}\rangle\langle 
\tilde{\psi}_{mn}|$ follow almost the same asymptotic behaviors. By leveraging the solution mentioned below Eq.~(\ref{real_2d}), we find that the rate $\lambda$ extracted from  $\langle\mathbf{r}'|P(E)|\mathbf{r}\rangle\propto e^{\lambda|\mathbf{r}'-\mathbf{r}|}\ (|\mathbf{r}'-\mathbf{r}|\gg 1)$ can be predicted as $\lambda=\max\{n_x\ln|\beta_x|+n_y\ln|\beta_y|;(\beta_x,\beta_y)\in L_E\}$, where $(n_x, n_y)=(\mathbf{r}'-\mathbf{r})/|\mathbf{r}'-\mathbf{r}|$ and $L_E$ is
the equienergy line on $\text{GBZ}_{2d}=\text{GBZ}_x\otimes\text{GBZ}_y$ that satisfies $h_x(k_x\rightarrow-i\ln\beta_x)+h_x(k_y\rightarrow-i\ln\beta_y)=E$. When $\text{GBZ}_x$ and $\text{GBZ}_x$ are the same non-circular GBZ, the rate $\lambda$ is positive for the both directions $(n_x,n_y)=(1,-1)/\sqrt{2}$ and $(n_x,n_y)=(-1,1)/\sqrt{2}$. This prediction is verified by the bi-directional amplification of $G_{2d}(\omega)$ shown in Fig.~\ref{figure_2d}(d). Furthermore, since the model considered here features non-reciprocal hoppings in real space, the bi-directional amplification coexists with unidirectional amplifications along $x$ and $y$ directions [Fig.~\ref{figure_2d}(e)]. 

As claimed before, the presence of V-shape Green’s function means that the spectrum is unstable under boundary perturbations. For the model of interest in this section, this is reflected in a real-to-complex spectral transition, which can be detected via the wavepacket motion governed by $i\partial_t |\psi(t)\rangle=H|\psi(t)\rangle$. The Fig.~\ref{figure_2d}(f) shows that the long-time behavior of the wavepacket evolving from the central of the system is significantly affected by the change on the boundary. The exponential growing rate of the wavefunction changes from zero to non-zero once the Hamiltonian $H_{2d}$ on the clean square is perturbed by adding $\delta B=\delta\sum_{\mathbf{r}\in \text{four corners}}|\mathbf{r}\rangle\langle\mathbf{r}|$. Moreover, the lower panel of Fig.~\ref{figure_2d}(f) shows that an exponentially small $\delta$ is strong enough to lead this transition. This is a unique phenomenon caused by V-shape Green's functions.

\begin{figure}
	\includegraphics[width=8cm]{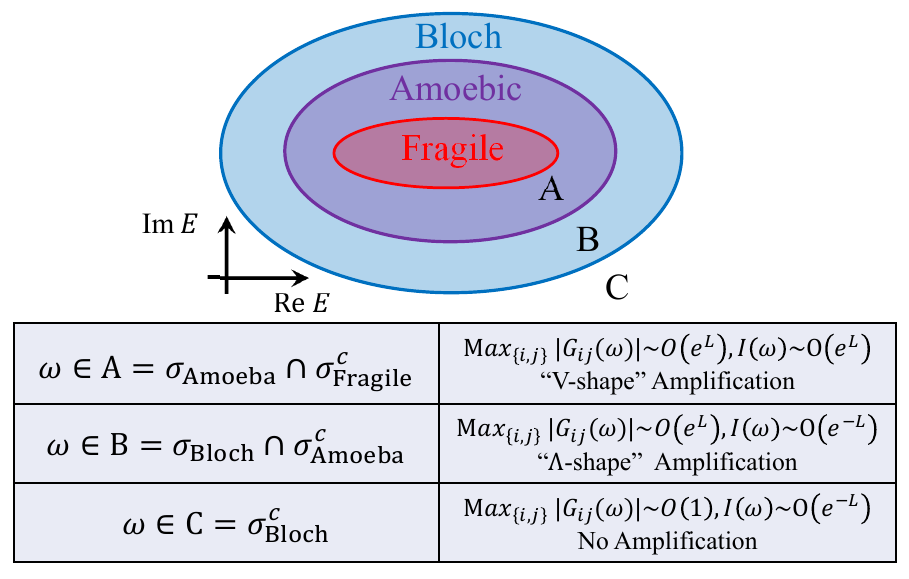}
\caption{Sketch of the hierarchy of non-Hermitian Green's functions. $\sigma_\alpha$ is the range on the complex plane inside the $\alpha$ spectrum, where $\alpha$ can be ``Bloch'', ``Amoeba'' or ``Fragile'', and the superscript $c$ denotes complement. $L$ denotes the length scale of the system.}\label{figure3}
\end{figure}
\emph{The hierarchical behaviors of non-Hermitian Green's functions.--} In this section, we discuss the general frequency range that allows the presence of V-shape Green’s function. To assist the analysis of $G(\omega)=(\omega-H)^{-1}$, we Hermitize $\omega-H$ as
\bea
H_\omega=\begin{pmatrix} 0 && \omega-H\\ \omega^*-H^\dag && 0\end{pmatrix}. \label{Hermitization}
\eea
It is easy to find that $G(\omega)$ is a block of the inverse of this Hamiltonian. Since $H_\omega$ is Hermitian, the majority of its eigenvalues, colloquially the bulk energies, can be predicted by the Bloch band theory. Moreover, the eigenvalues of $H_\omega$ also possibly contain exponentially small zero modes that stem from the non-trivial band topology and the specialty of boundary conditions. We term them as topological zero modes and fragile zero modes, respectively. For the $H_\omega$ with only the chiral symmetry requested by its structure, topological zero modes are related to the homotopy class of $H_\omega$.  This is characterized by the invariant $W_i=\int_{[0,2\pi]^d} d^d k/(2\pi)^d i\partial_{k_i} \ln \det[\omega-h(\mathbf{k})]$ $(i=x,y,\ldots)$ in $d$ dimensions, where $h(\mathbf{k})$ marks the Bloch Hamiltonian corresponding to $H$. A crucial observation is that $W_i$ can be tuned to zero by the analytic extension $h(\mathbf{k}\rightarrow \mathbf{k}-i\boldsymbol{\mu}$) with a proper $\boldsymbol{\mu}$. This means that only bulk modes and fragile zero modes survive in the spectrum of the Hamiltonian $H'_\omega$ given by the Hermitization of $\omega-T_{\boldsymbol{\mu}}^{-1} HT_{\boldsymbol{\mu}}$  with $T_{\boldsymbol{\mu}}$ following $T_{\boldsymbol{\mu}}|\mathbf{r}\rangle=e^{{\boldsymbol\mu}\cdot\mathbf{r}}|\mathbf{r}\rangle$. Exploiting the facts that the contribution of the bulk modes to $(H'_\omega)^{-1}$ can be expressed as an integral along the BZ and $G(\omega)$ is a block of $T_{\boldsymbol{\mu}}(H'_\omega)^{-1}T_{\boldsymbol{\mu}}^{-1}$, one can readily derive
\bea
\langle \mathbf{r} |G(\omega)|\mathbf{r'}
\rangle=\int_{[0,2\pi]^d} \frac{d^d k}{(2\pi)^d} \frac{e^{i(\mathbf{k}-i{\boldsymbol{\mu})\cdot(\mathbf{r}-\mathbf{r'})}}}{\omega-h(\mathbf{k}-i\boldsymbol{\mu})}+G_f(\omega). \label{Green}
\eea
The $G_f(\omega)$ in the above formula is brought by the fragile zero modes of $H'_\omega$. If $G_f(\omega)=0$, one can prove that $|\langle \mathbf{r} |G(\omega)|\mathbf{r'}
\rangle| \leq e^{{\boldsymbol{\mu}}\cdot(\mathbf{r}-\mathbf{r'})} \int_{[0,2\pi]^d} d^d k/(2\pi)^d |[\omega-h(\mathbf{k}-i\boldsymbol{\mu})]^{-1}|$, which excludes $G(\omega)$ from being V-shape. Consequently, a necessary condition for $G(\omega)$ to be V-shape is the existence of fragile zero modes in the spectrum of $H'_\omega$. Thus, fragile zero modes also result in the mismatch between the amoebic spectrum and the fragile spectrum under a specific geometry \cite{wang2024singular,SM}. Following this understanding, we have proved that any fragile spectrum is always enclosed by the amoebic spectrum \cite{wang2024singular,SM}. As explained in SM \cite{SM}, this property often indicates $G_f(\omega)=0$ and thus forbids the V-shape behaviors of $G(\omega)$ for the frequency $\omega$ outside the amoebic spectrum. Additionally, for the frequency $\omega$ also outside the Bloch spectrum, $G_f(\omega)=0$ is satisfied along with $\boldsymbol{\mu}=0$ in Eq.~(\ref{Green}). Thus, the Green's function with such frequencies cannot be exponentially amplified. To summarize, the asymptotic behaviors of non-Hermitian Green's functions generally follow the hierarchy shown in Fig.~\ref{figure3}, where the table lists the allowed behavior of $G(\omega)$ for the frequency $\omega$ within three ranges. This hierarchy is obeyed by the aforementioned examples.

\emph{Conclusions.--} Our work unveils that the spectral fragility in non-Hermitian bands can be quantitatively characterized by unconventional V-shape asymptotic behaviors in their Green's functions. This result provides a valuable perspective for understanding how the fragile spectrum transmutes into the stable (i.e., amoebic) spectrum upon generic local perturbations. Furthermore, we uncover how the asymptotic behaviors of a non-Hermitian Green's function vary hierarchically in terms of its frequency. This hierarchy indicates that a prerequisite for producing V-shape Green's functions is the mismatch between the spectrum under a regular geometry and the amoebic spectrum. This mismatch widely exists in two and higher dimensions \cite{wang2024singular,xu2023two}, thereby ensuring the ubiquity of V-shape Green's functions. Our predictions can be verified on the state-of-the-art platforms, including topoelectrical circuits \cite{helbig2020generalized}, active mechanical systems \cite{ghatak2020observation} and photonics \cite{xiao2024burst,lin2023manipulating,xue2024self}.



\emph{Acknowledgments.--} This work is supported by NSFC under Grant No.~12125405 and Grant No.~12404189. FS also acknowledges support from the Postdoctoral Fellowship Program of CPSF under Grant No. GZB20240732.

\emph{Note added.--} Upon finalizing this draft, we became aware of a recent preprint \cite{shu2024ultra} with partial overlap with our work.  


\bibliography{references}

\begin{thebibliography}{71}%
\makeatletter
\providecommand \@ifxundefined [1]{%
 \@ifx{#1\undefined}
}%
\providecommand \@ifnum [1]{%
 \ifnum #1\expandafter \@firstoftwo
 \else \expandafter \@secondoftwo
 \fi
}%
\providecommand \@ifx [1]{%
 \ifx #1\expandafter \@firstoftwo
 \else \expandafter \@secondoftwo
 \fi
}%
\providecommand \natexlab [1]{#1}%
\providecommand \enquote  [1]{``#1''}%
\providecommand \bibnamefont  [1]{#1}%
\providecommand \bibfnamefont [1]{#1}%
\providecommand \citenamefont [1]{#1}%
\providecommand \href@noop [0]{\@secondoftwo}%
\providecommand \href [0]{\begingroup \@sanitize@url \@href}%
\providecommand \@href[1]{\@@startlink{#1}\@@href}%
\providecommand \@@href[1]{\endgroup#1\@@endlink}%
\providecommand \@sanitize@url [0]{\catcode `\\12\catcode `\$12\catcode `\&12\catcode `\#12\catcode `\^12\catcode `\_12\catcode `\%12\relax}%
\providecommand \@@startlink[1]{}%
\providecommand \@@endlink[0]{}%
\providecommand \url  [0]{\begingroup\@sanitize@url \@url }%
\providecommand \@url [1]{\endgroup\@href {#1}{\urlprefix }}%
\providecommand \urlprefix  [0]{URL }%
\providecommand \Eprint [0]{\href }%
\providecommand \doibase [0]{https://doi.org/}%
\providecommand \selectlanguage [0]{\@gobble}%
\providecommand \bibinfo  [0]{\@secondoftwo}%
\providecommand \bibfield  [0]{\@secondoftwo}%
\providecommand \translation [1]{[#1]}%
\providecommand \BibitemOpen [0]{}%
\providecommand \bibitemStop [0]{}%
\providecommand \bibitemNoStop [0]{.\EOS\space}%
\providecommand \EOS [0]{\spacefactor3000\relax}%
\providecommand \BibitemShut  [1]{\csname bibitem#1\endcsname}%
\let\auto@bib@innerbib\@empty
\bibitem [{\citenamefont {Ashida}\ \emph {et~al.}(2020)\citenamefont {Ashida}, \citenamefont {Gong},\ and\ \citenamefont {Ueda}}]{ashida2020nonhermitian}%
  \BibitemOpen
  \bibfield  {author} {\bibinfo {author} {\bibfnamefont {Y.}~\bibnamefont {Ashida}}, \bibinfo {author} {\bibfnamefont {Z.}~\bibnamefont {Gong}},\ and\ \bibinfo {author} {\bibfnamefont {M.}~\bibnamefont {Ueda}},\ }\bibfield  {title} {\bibinfo {title} {Non-{{Hermitian}} physics},\ }\href {https://doi.org/10.1080/00018732.2021.1876991} {\bibfield  {journal} {\bibinfo  {journal} {Advances in Physics}\ }\textbf {\bibinfo {volume} {69}},\ \bibinfo {pages} {249} (\bibinfo {year} {2020})}\BibitemShut {NoStop}%
\bibitem [{\citenamefont {Bergholtz}\ \emph {et~al.}(2021)\citenamefont {Bergholtz}, \citenamefont {Budich},\ and\ \citenamefont {Kunst}}]{bergholtz2021exceptional}%
  \BibitemOpen
  \bibfield  {author} {\bibinfo {author} {\bibfnamefont {E.~J.}\ \bibnamefont {Bergholtz}}, \bibinfo {author} {\bibfnamefont {J.~C.}\ \bibnamefont {Budich}},\ and\ \bibinfo {author} {\bibfnamefont {F.~K.}\ \bibnamefont {Kunst}},\ }\bibfield  {title} {\bibinfo {title} {Exceptional topology of non-{{Hermitian}} systems},\ }\href {https://doi.org/10.1103/RevModPhys.93.015005} {\bibfield  {journal} {\bibinfo  {journal} {Rev. Mod. Phys.}\ }\textbf {\bibinfo {volume} {93}},\ \bibinfo {pages} {015005} (\bibinfo {year} {2021})}\BibitemShut {NoStop}%
\bibitem [{\citenamefont {Yao}\ and\ \citenamefont {Wang}(2018)}]{yao2018edge}%
  \BibitemOpen
  \bibfield  {author} {\bibinfo {author} {\bibfnamefont {S.}~\bibnamefont {Yao}}\ and\ \bibinfo {author} {\bibfnamefont {Z.}~\bibnamefont {Wang}},\ }\bibfield  {title} {\bibinfo {title} {Edge {{States}} and {{Topological Invariants}} of {{Non-Hermitian Systems}}},\ }\href {https://doi.org/10.1103/PhysRevLett.121.086803} {\bibfield  {journal} {\bibinfo  {journal} {Phys. Rev. Lett.}\ }\textbf {\bibinfo {volume} {121}},\ \bibinfo {pages} {086803} (\bibinfo {year} {2018})}\BibitemShut {NoStop}%
\bibitem [{\citenamefont {Yao}\ \emph {et~al.}(2018)\citenamefont {Yao}, \citenamefont {Song},\ and\ \citenamefont {Wang}}]{yao2018nonhermitian}%
  \BibitemOpen
  \bibfield  {author} {\bibinfo {author} {\bibfnamefont {S.}~\bibnamefont {Yao}}, \bibinfo {author} {\bibfnamefont {F.}~\bibnamefont {Song}},\ and\ \bibinfo {author} {\bibfnamefont {Z.}~\bibnamefont {Wang}},\ }\bibfield  {title} {\bibinfo {title} {Non-{{Hermitian Chern Bands}}},\ }\href {https://doi.org/10.1103/PhysRevLett.121.136802} {\bibfield  {journal} {\bibinfo  {journal} {Phys. Rev. Lett.}\ }\textbf {\bibinfo {volume} {121}},\ \bibinfo {pages} {136802} (\bibinfo {year} {2018})}\BibitemShut {NoStop}%
\bibitem [{\citenamefont {Kunst}\ \emph {et~al.}(2018)\citenamefont {Kunst}, \citenamefont {Edvardsson}, \citenamefont {Budich},\ and\ \citenamefont {Bergholtz}}]{kunst2018biorthogonal}%
  \BibitemOpen
  \bibfield  {author} {\bibinfo {author} {\bibfnamefont {F.~K.}\ \bibnamefont {Kunst}}, \bibinfo {author} {\bibfnamefont {E.}~\bibnamefont {Edvardsson}}, \bibinfo {author} {\bibfnamefont {J.~C.}\ \bibnamefont {Budich}},\ and\ \bibinfo {author} {\bibfnamefont {E.~J.}\ \bibnamefont {Bergholtz}},\ }\bibfield  {title} {\bibinfo {title} {Biorthogonal {{Bulk-Boundary Correspondence}} in {{Non-Hermitian Systems}}},\ }\href {https://doi.org/10.1103/PhysRevLett.121.026808} {\bibfield  {journal} {\bibinfo  {journal} {Phys. Rev. Lett.}\ }\textbf {\bibinfo {volume} {121}},\ \bibinfo {pages} {026808} (\bibinfo {year} {2018})}\BibitemShut {NoStop}%
\bibitem [{\citenamefont {Lee}\ and\ \citenamefont {Thomale}(2019)}]{lee2019anatomy}%
  \BibitemOpen
  \bibfield  {author} {\bibinfo {author} {\bibfnamefont {C.~H.}\ \bibnamefont {Lee}}\ and\ \bibinfo {author} {\bibfnamefont {R.}~\bibnamefont {Thomale}},\ }\bibfield  {title} {\bibinfo {title} {Anatomy of skin modes and topology in non-{{Hermitian}} systems},\ }\href {https://doi.org/10.1103/PhysRevB.99.201103} {\bibfield  {journal} {\bibinfo  {journal} {Phys. Rev. B}\ }\textbf {\bibinfo {volume} {99}},\ \bibinfo {pages} {201103} (\bibinfo {year} {2019})}\BibitemShut {NoStop}%
\bibitem [{\citenamefont {Martinez~Alvarez}\ \emph {et~al.}(2018)\citenamefont {Martinez~Alvarez}, \citenamefont {Barrios~Vargas},\ and\ \citenamefont {Foa~Torres}}]{martinezalvarez2018nonhermitian}%
  \BibitemOpen
  \bibfield  {author} {\bibinfo {author} {\bibfnamefont {V.~M.}\ \bibnamefont {Martinez~Alvarez}}, \bibinfo {author} {\bibfnamefont {J.~E.}\ \bibnamefont {Barrios~Vargas}},\ and\ \bibinfo {author} {\bibfnamefont {L.~E.~F.}\ \bibnamefont {Foa~Torres}},\ }\bibfield  {title} {\bibinfo {title} {Non-{{Hermitian}} robust edge states in one dimension: {{Anomalous}} localization and eigenspace condensation at exceptional points},\ }\href {https://doi.org/10.1103/PhysRevB.97.121401} {\bibfield  {journal} {\bibinfo  {journal} {Phys. Rev. B}\ }\textbf {\bibinfo {volume} {97}},\ \bibinfo {pages} {121401} (\bibinfo {year} {2018})}\BibitemShut {NoStop}%
\bibitem [{\citenamefont {Ghatak}\ \emph {et~al.}(2020)\citenamefont {Ghatak}, \citenamefont {Brandenbourger}, \citenamefont {Van~Wezel},\ and\ \citenamefont {Coulais}}]{ghatak2020observation}%
  \BibitemOpen
  \bibfield  {author} {\bibinfo {author} {\bibfnamefont {A.}~\bibnamefont {Ghatak}}, \bibinfo {author} {\bibfnamefont {M.}~\bibnamefont {Brandenbourger}}, \bibinfo {author} {\bibfnamefont {J.}~\bibnamefont {Van~Wezel}},\ and\ \bibinfo {author} {\bibfnamefont {C.}~\bibnamefont {Coulais}},\ }\bibfield  {title} {\bibinfo {title} {Observation of non-{{Hermitian}} topology and its bulk–edge correspondence in an active mechanical metamaterial},\ }\href {https://doi.org/10.1073/pnas.2010580117} {\bibfield  {journal} {\bibinfo  {journal} {Proc. Natl. Acad. Sci. U.S.A.}\ }\textbf {\bibinfo {volume} {117}},\ \bibinfo {pages} {29561} (\bibinfo {year} {2020})}\BibitemShut {NoStop}%
\bibitem [{\citenamefont {Xiao}\ \emph {et~al.}(2020)\citenamefont {Xiao}, \citenamefont {Deng}, \citenamefont {Wang}, \citenamefont {Zhu}, \citenamefont {Wang}, \citenamefont {Yi},\ and\ \citenamefont {Xue}}]{xiao2020nonhermitian}%
  \BibitemOpen
  \bibfield  {author} {\bibinfo {author} {\bibfnamefont {L.}~\bibnamefont {Xiao}}, \bibinfo {author} {\bibfnamefont {T.}~\bibnamefont {Deng}}, \bibinfo {author} {\bibfnamefont {K.}~\bibnamefont {Wang}}, \bibinfo {author} {\bibfnamefont {G.}~\bibnamefont {Zhu}}, \bibinfo {author} {\bibfnamefont {Z.}~\bibnamefont {Wang}}, \bibinfo {author} {\bibfnamefont {W.}~\bibnamefont {Yi}},\ and\ \bibinfo {author} {\bibfnamefont {P.}~\bibnamefont {Xue}},\ }\bibfield  {title} {\bibinfo {title} {Non-{{Hermitian}} bulk–boundary correspondence in quantum dynamics},\ }\href {https://doi.org/10.1038/s41567-020-0836-6} {\bibfield  {journal} {\bibinfo  {journal} {Nat. Phys.}\ }\textbf {\bibinfo {volume} {16}},\ \bibinfo {pages} {761} (\bibinfo {year} {2020})}\BibitemShut {NoStop}%
\bibitem [{\citenamefont {Helbig}\ \emph {et~al.}(2020)\citenamefont {Helbig}, \citenamefont {Hofmann}, \citenamefont {Imhof}, \citenamefont {Abdelghany}, \citenamefont {Kiessling}, \citenamefont {Molenkamp}, \citenamefont {Lee}, \citenamefont {Szameit}, \citenamefont {Greiter},\ and\ \citenamefont {Thomale}}]{helbig2020generalized}%
  \BibitemOpen
  \bibfield  {author} {\bibinfo {author} {\bibfnamefont {T.}~\bibnamefont {Helbig}}, \bibinfo {author} {\bibfnamefont {T.}~\bibnamefont {Hofmann}}, \bibinfo {author} {\bibfnamefont {S.}~\bibnamefont {Imhof}}, \bibinfo {author} {\bibfnamefont {M.}~\bibnamefont {Abdelghany}}, \bibinfo {author} {\bibfnamefont {T.}~\bibnamefont {Kiessling}}, \bibinfo {author} {\bibfnamefont {L.~W.}\ \bibnamefont {Molenkamp}}, \bibinfo {author} {\bibfnamefont {C.~H.}\ \bibnamefont {Lee}}, \bibinfo {author} {\bibfnamefont {A.}~\bibnamefont {Szameit}}, \bibinfo {author} {\bibfnamefont {M.}~\bibnamefont {Greiter}},\ and\ \bibinfo {author} {\bibfnamefont {R.}~\bibnamefont {Thomale}},\ }\bibfield  {title} {\bibinfo {title} {Generalized bulk–boundary correspondence in non-{{Hermitian}} topolectrical circuits},\ }\href {https://doi.org/10.1038/s41567-020-0922-9} {\bibfield  {journal} {\bibinfo  {journal} {Nat. Phys.}\ }\textbf {\bibinfo {volume} {16}},\ \bibinfo {pages} {747} (\bibinfo {year} {2020})}\BibitemShut {NoStop}%
\bibitem [{\citenamefont {Weidemann}\ \emph {et~al.}(2020)\citenamefont {Weidemann}, \citenamefont {Kremer}, \citenamefont {Helbig}, \citenamefont {Hofmann}, \citenamefont {Stegmaier}, \citenamefont {Greiter}, \citenamefont {Thomale},\ and\ \citenamefont {Szameit}}]{weidemann2020topological}%
  \BibitemOpen
  \bibfield  {author} {\bibinfo {author} {\bibfnamefont {S.}~\bibnamefont {Weidemann}}, \bibinfo {author} {\bibfnamefont {M.}~\bibnamefont {Kremer}}, \bibinfo {author} {\bibfnamefont {T.}~\bibnamefont {Helbig}}, \bibinfo {author} {\bibfnamefont {T.}~\bibnamefont {Hofmann}}, \bibinfo {author} {\bibfnamefont {A.}~\bibnamefont {Stegmaier}}, \bibinfo {author} {\bibfnamefont {M.}~\bibnamefont {Greiter}}, \bibinfo {author} {\bibfnamefont {R.}~\bibnamefont {Thomale}},\ and\ \bibinfo {author} {\bibfnamefont {A.}~\bibnamefont {Szameit}},\ }\bibfield  {title} {\bibinfo {title} {Topological funneling of light},\ }\href {https://doi.org/10.1126/science.aaz8727} {\bibfield  {journal} {\bibinfo  {journal} {Science}\ }\textbf {\bibinfo {volume} {368}},\ \bibinfo {pages} {311} (\bibinfo {year} {2020})}\BibitemShut {NoStop}%
\bibitem [{\citenamefont {Wang}\ \emph {et~al.}(2022{\natexlab{a}})\citenamefont {Wang}, \citenamefont {Wang},\ and\ \citenamefont {Ma}}]{wang2022nonhermitian}%
  \BibitemOpen
  \bibfield  {author} {\bibinfo {author} {\bibfnamefont {W.}~\bibnamefont {Wang}}, \bibinfo {author} {\bibfnamefont {X.}~\bibnamefont {Wang}},\ and\ \bibinfo {author} {\bibfnamefont {G.}~\bibnamefont {Ma}},\ }\bibfield  {title} {\bibinfo {title} {Non-{{Hermitian}} morphing of topological modes},\ }\href {https://doi.org/10.1038/s41586-022-04929-1} {\bibfield  {journal} {\bibinfo  {journal} {Nature}\ }\textbf {\bibinfo {volume} {608}},\ \bibinfo {pages} {50} (\bibinfo {year} {2022}{\natexlab{a}})}\BibitemShut {NoStop}%
\bibitem [{\citenamefont {Liang}\ \emph {et~al.}(2022)\citenamefont {Liang}, \citenamefont {Xie}, \citenamefont {Dong}, \citenamefont {Li}, \citenamefont {Li}, \citenamefont {Gadway}, \citenamefont {Yi},\ and\ \citenamefont {Yan}}]{liang2022dynamic}%
  \BibitemOpen
  \bibfield  {author} {\bibinfo {author} {\bibfnamefont {Q.}~\bibnamefont {Liang}}, \bibinfo {author} {\bibfnamefont {D.}~\bibnamefont {Xie}}, \bibinfo {author} {\bibfnamefont {Z.}~\bibnamefont {Dong}}, \bibinfo {author} {\bibfnamefont {H.}~\bibnamefont {Li}}, \bibinfo {author} {\bibfnamefont {H.}~\bibnamefont {Li}}, \bibinfo {author} {\bibfnamefont {B.}~\bibnamefont {Gadway}}, \bibinfo {author} {\bibfnamefont {W.}~\bibnamefont {Yi}},\ and\ \bibinfo {author} {\bibfnamefont {B.}~\bibnamefont {Yan}},\ }\bibfield  {title} {\bibinfo {title} {Dynamic {{Signatures}} of {{Non-Hermitian Skin Effect}} and {{Topology}} in {{Ultracold Atoms}}},\ }\href {https://doi.org/10.1103/PhysRevLett.129.070401} {\bibfield  {journal} {\bibinfo  {journal} {Phys. Rev. Lett.}\ }\textbf {\bibinfo {volume} {129}},\ \bibinfo {pages} {070401} (\bibinfo {year} {2022})}\BibitemShut {NoStop}%
\bibitem [{\citenamefont {Zhang}\ \emph {et~al.}(2022{\natexlab{a}})\citenamefont {Zhang}, \citenamefont {Zhang}, \citenamefont {Lu},\ and\ \citenamefont {Chen}}]{zhang2022review}%
  \BibitemOpen
  \bibfield  {author} {\bibinfo {author} {\bibfnamefont {X.}~\bibnamefont {Zhang}}, \bibinfo {author} {\bibfnamefont {T.}~\bibnamefont {Zhang}}, \bibinfo {author} {\bibfnamefont {M.-H.}\ \bibnamefont {Lu}},\ and\ \bibinfo {author} {\bibfnamefont {Y.-F.}\ \bibnamefont {Chen}},\ }\bibfield  {title} {\bibinfo {title} {A review on non-{{Hermitian}} skin effect},\ }\href {https://doi.org/10.1080/23746149.2022.2109431} {\bibfield  {journal} {\bibinfo  {journal} {Advances in Physics: X}\ }\textbf {\bibinfo {volume} {7}},\ \bibinfo {pages} {2109431} (\bibinfo {year} {2022}{\natexlab{a}})}\BibitemShut {NoStop}%
\bibitem [{\citenamefont {Ding}\ \emph {et~al.}(2022)\citenamefont {Ding}, \citenamefont {Fang},\ and\ \citenamefont {Ma}}]{ding2022nonhermitian}%
  \BibitemOpen
  \bibfield  {author} {\bibinfo {author} {\bibfnamefont {K.}~\bibnamefont {Ding}}, \bibinfo {author} {\bibfnamefont {C.}~\bibnamefont {Fang}},\ and\ \bibinfo {author} {\bibfnamefont {G.}~\bibnamefont {Ma}},\ }\bibfield  {title} {\bibinfo {title} {Non-{{Hermitian}} topology and exceptional-point geometries},\ }\href {https://doi.org/10.1038/s42254-022-00516-5} {\bibfield  {journal} {\bibinfo  {journal} {Nat Rev Phys}\ }\textbf {\bibinfo {volume} {4}},\ \bibinfo {pages} {745} (\bibinfo {year} {2022})}\BibitemShut {NoStop}%
\bibitem [{\citenamefont {Song}\ \emph {et~al.}(2019)\citenamefont {Song}, \citenamefont {Yao},\ and\ \citenamefont {Wang}}]{song2019nonhermitian}%
  \BibitemOpen
  \bibfield  {author} {\bibinfo {author} {\bibfnamefont {F.}~\bibnamefont {Song}}, \bibinfo {author} {\bibfnamefont {S.}~\bibnamefont {Yao}},\ and\ \bibinfo {author} {\bibfnamefont {Z.}~\bibnamefont {Wang}},\ }\bibfield  {title} {\bibinfo {title} {Non-{{Hermitian Skin Effect}} and {{Chiral Damping}} in {{Open Quantum Systems}}},\ }\href {https://doi.org/10.1103/PhysRevLett.123.170401} {\bibfield  {journal} {\bibinfo  {journal} {Phys. Rev. Lett.}\ }\textbf {\bibinfo {volume} {123}},\ \bibinfo {pages} {170401} (\bibinfo {year} {2019})}\BibitemShut {NoStop}%
\bibitem [{\citenamefont {Haga}\ \emph {et~al.}(2021)\citenamefont {Haga}, \citenamefont {Nakagawa}, \citenamefont {Hamazaki},\ and\ \citenamefont {Ueda}}]{haga2021liouvillian}%
  \BibitemOpen
  \bibfield  {author} {\bibinfo {author} {\bibfnamefont {T.}~\bibnamefont {Haga}}, \bibinfo {author} {\bibfnamefont {M.}~\bibnamefont {Nakagawa}}, \bibinfo {author} {\bibfnamefont {R.}~\bibnamefont {Hamazaki}},\ and\ \bibinfo {author} {\bibfnamefont {M.}~\bibnamefont {Ueda}},\ }\bibfield  {title} {\bibinfo {title} {Liouvillian {{Skin Effect}}: {{Slowing Down}} of {{Relaxation Processes}} without {{Gap Closing}}},\ }\href {https://doi.org/10.1103/PhysRevLett.127.070402} {\bibfield  {journal} {\bibinfo  {journal} {Phys. Rev. Lett.}\ }\textbf {\bibinfo {volume} {127}},\ \bibinfo {pages} {070402} (\bibinfo {year} {2021})}\BibitemShut {NoStop}%
\bibitem [{\citenamefont {Liu}\ \emph {et~al.}(2020)\citenamefont {Liu}, \citenamefont {Zhang}, \citenamefont {Yang},\ and\ \citenamefont {Chen}}]{liu2020helical}%
  \BibitemOpen
  \bibfield  {author} {\bibinfo {author} {\bibfnamefont {C.-H.}\ \bibnamefont {Liu}}, \bibinfo {author} {\bibfnamefont {K.}~\bibnamefont {Zhang}}, \bibinfo {author} {\bibfnamefont {Z.}~\bibnamefont {Yang}},\ and\ \bibinfo {author} {\bibfnamefont {S.}~\bibnamefont {Chen}},\ }\bibfield  {title} {\bibinfo {title} {Helical damping and dynamical critical skin effect in open quantum systems},\ }\href {https://doi.org/10.1103/PhysRevResearch.2.043167} {\bibfield  {journal} {\bibinfo  {journal} {Phys. Rev. Research}\ }\textbf {\bibinfo {volume} {2}},\ \bibinfo {pages} {043167} (\bibinfo {year} {2020})}\BibitemShut {NoStop}%
\bibitem [{\citenamefont {Yang}\ \emph {et~al.}(2022)\citenamefont {Yang}, \citenamefont {Jiang},\ and\ \citenamefont {Bergholtz}}]{yang2022liouvillian}%
  \BibitemOpen
  \bibfield  {author} {\bibinfo {author} {\bibfnamefont {F.}~\bibnamefont {Yang}}, \bibinfo {author} {\bibfnamefont {Q.-D.}\ \bibnamefont {Jiang}},\ and\ \bibinfo {author} {\bibfnamefont {E.~J.}\ \bibnamefont {Bergholtz}},\ }\bibfield  {title} {\bibinfo {title} {Liouvillian skin effect in an exactly solvable model},\ }\href {https://doi.org/10.1103/PhysRevResearch.4.023160} {\bibfield  {journal} {\bibinfo  {journal} {Phys. Rev. Research}\ }\textbf {\bibinfo {volume} {4}},\ \bibinfo {pages} {023160} (\bibinfo {year} {2022})}\BibitemShut {NoStop}%
\bibitem [{\citenamefont {Xue}\ \emph {et~al.}(2022)\citenamefont {Xue}, \citenamefont {Hu}, \citenamefont {Song},\ and\ \citenamefont {Wang}}]{xue2022nonhermitian}%
  \BibitemOpen
  \bibfield  {author} {\bibinfo {author} {\bibfnamefont {W.-T.}\ \bibnamefont {Xue}}, \bibinfo {author} {\bibfnamefont {Y.-M.}\ \bibnamefont {Hu}}, \bibinfo {author} {\bibfnamefont {F.}~\bibnamefont {Song}},\ and\ \bibinfo {author} {\bibfnamefont {Z.}~\bibnamefont {Wang}},\ }\bibfield  {title} {\bibinfo {title} {Non-{{Hermitian Edge Burst}}},\ }\href {https://doi.org/10.1103/PhysRevLett.128.120401} {\bibfield  {journal} {\bibinfo  {journal} {Phys. Rev. Lett.}\ }\textbf {\bibinfo {volume} {128}},\ \bibinfo {pages} {120401} (\bibinfo {year} {2022})}\BibitemShut {NoStop}%
\bibitem [{\citenamefont {Wang}\ \emph {et~al.}(2021)\citenamefont {Wang}, \citenamefont {Liu},\ and\ \citenamefont {Zhang}}]{wang2021quantum}%
  \BibitemOpen
  \bibfield  {author} {\bibinfo {author} {\bibfnamefont {L.}~\bibnamefont {Wang}}, \bibinfo {author} {\bibfnamefont {Q.}~\bibnamefont {Liu}},\ and\ \bibinfo {author} {\bibfnamefont {Y.}~\bibnamefont {Zhang}},\ }\bibfield  {title} {\bibinfo {title} {Quantum dynamics on a lossy non-{{Hermitian}} lattice*},\ }\href {https://doi.org/10.1088/1674-1056/abd765} {\bibfield  {journal} {\bibinfo  {journal} {Chinese Phys. B}\ }\textbf {\bibinfo {volume} {30}},\ \bibinfo {pages} {020506} (\bibinfo {year} {2021})}\BibitemShut {NoStop}%
\bibitem [{\citenamefont {Hu}\ \emph {et~al.}(2023{\natexlab{a}})\citenamefont {Hu}, \citenamefont {Xue}, \citenamefont {Song},\ and\ \citenamefont {Wang}}]{Hu2023burst}%
  \BibitemOpen
  \bibfield  {author} {\bibinfo {author} {\bibfnamefont {Y.-M.}\ \bibnamefont {Hu}}, \bibinfo {author} {\bibfnamefont {W.-T.}\ \bibnamefont {Xue}}, \bibinfo {author} {\bibfnamefont {F.}~\bibnamefont {Song}},\ and\ \bibinfo {author} {\bibfnamefont {Z.}~\bibnamefont {Wang}},\ }\bibfield  {title} {\bibinfo {title} {Steady-state edge burst: From free-particle systems to interaction-induced phenomena},\ }\href {https://doi.org/10.1103/PhysRevB.108.235422} {\bibfield  {journal} {\bibinfo  {journal} {Phys. Rev. B}\ }\textbf {\bibinfo {volume} {108}},\ \bibinfo {pages} {235422} (\bibinfo {year} {2023}{\natexlab{a}})}\BibitemShut {NoStop}%
\bibitem [{\citenamefont {Xiao}\ \emph {et~al.}(2024)\citenamefont {Xiao}, \citenamefont {Xue}, \citenamefont {Song}, \citenamefont {Hu}, \citenamefont {Yi}, \citenamefont {Wang},\ and\ \citenamefont {Xue}}]{xiao2024burst}%
  \BibitemOpen
  \bibfield  {author} {\bibinfo {author} {\bibfnamefont {L.}~\bibnamefont {Xiao}}, \bibinfo {author} {\bibfnamefont {W.-T.}\ \bibnamefont {Xue}}, \bibinfo {author} {\bibfnamefont {F.}~\bibnamefont {Song}}, \bibinfo {author} {\bibfnamefont {Y.-M.}\ \bibnamefont {Hu}}, \bibinfo {author} {\bibfnamefont {W.}~\bibnamefont {Yi}}, \bibinfo {author} {\bibfnamefont {Z.}~\bibnamefont {Wang}},\ and\ \bibinfo {author} {\bibfnamefont {P.}~\bibnamefont {Xue}},\ }\bibfield  {title} {\bibinfo {title} {Observation of non-hermitian edge burst in quantum dynamics},\ }\href {https://doi.org/10.1103/PhysRevLett.133.070801} {\bibfield  {journal} {\bibinfo  {journal} {Phys. Rev. Lett.}\ }\textbf {\bibinfo {volume} {133}},\ \bibinfo {pages} {070801} (\bibinfo {year} {2024})}\BibitemShut {NoStop}%
\bibitem [{\citenamefont {Zhu}\ \emph {et~al.}(2024)\citenamefont {Zhu}, \citenamefont {Mao}, \citenamefont {Chen}, \citenamefont {Yang}, \citenamefont {Li}, \citenamefont {Yang}, \citenamefont {Li},\ and\ \citenamefont {Fan}}]{zhu2024burst}%
  \BibitemOpen
  \bibfield  {author} {\bibinfo {author} {\bibfnamefont {J.}~\bibnamefont {Zhu}}, \bibinfo {author} {\bibfnamefont {Y.-L.}\ \bibnamefont {Mao}}, \bibinfo {author} {\bibfnamefont {H.}~\bibnamefont {Chen}}, \bibinfo {author} {\bibfnamefont {K.-X.}\ \bibnamefont {Yang}}, \bibinfo {author} {\bibfnamefont {L.}~\bibnamefont {Li}}, \bibinfo {author} {\bibfnamefont {B.}~\bibnamefont {Yang}}, \bibinfo {author} {\bibfnamefont {Z.-D.}\ \bibnamefont {Li}},\ and\ \bibinfo {author} {\bibfnamefont {J.}~\bibnamefont {Fan}},\ }\bibfield  {title} {\bibinfo {title} {Observation of non-hermitian edge burst effect in one-dimensional photonic quantum walk},\ }\href {https://doi.org/10.1103/PhysRevLett.132.203801} {\bibfield  {journal} {\bibinfo  {journal} {Phys. Rev. Lett.}\ }\textbf {\bibinfo {volume} {132}},\ \bibinfo {pages} {203801} (\bibinfo {year} {2024})}\BibitemShut {NoStop}%
\bibitem [{\citenamefont {Yi}\ and\ \citenamefont {Yang}(2020)}]{yi2020nonhermitian}%
  \BibitemOpen
  \bibfield  {author} {\bibinfo {author} {\bibfnamefont {Y.}~\bibnamefont {Yi}}\ and\ \bibinfo {author} {\bibfnamefont {Z.}~\bibnamefont {Yang}},\ }\bibfield  {title} {\bibinfo {title} {Non-{{Hermitian Skin Modes Induced}} by {{On-Site Dissipations}} and {{Chiral Tunneling Effect}}},\ }\href {https://doi.org/10.1103/PhysRevLett.125.186802} {\bibfield  {journal} {\bibinfo  {journal} {Phys. Rev. Lett.}\ }\textbf {\bibinfo {volume} {125}},\ \bibinfo {pages} {186802} (\bibinfo {year} {2020})}\BibitemShut {NoStop}%
\bibitem [{\citenamefont {Longhi}(2020)}]{longhi2020nonblochband}%
  \BibitemOpen
  \bibfield  {author} {\bibinfo {author} {\bibfnamefont {S.}~\bibnamefont {Longhi}},\ }\bibfield  {title} {\bibinfo {title} {Non-{{Bloch-Band Collapse}} and {{Chiral Zener Tunneling}}},\ }\href {https://doi.org/10.1103/PhysRevLett.124.066602} {\bibfield  {journal} {\bibinfo  {journal} {Phys. Rev. Lett.}\ }\textbf {\bibinfo {volume} {124}},\ \bibinfo {pages} {066602} (\bibinfo {year} {2020})}\BibitemShut {NoStop}%
\bibitem [{\citenamefont {Longhi}(2019{\natexlab{a}})}]{longhi2019nonbloch}%
  \BibitemOpen
  \bibfield  {author} {\bibinfo {author} {\bibfnamefont {S.}~\bibnamefont {Longhi}},\ }\bibfield  {title} {\bibinfo {title} {Non-{{Bloch}} $\mathcal{PT}$ symmetry breaking in non-{{Hermitian}} photonic quantum walks},\ }\href {https://doi.org/10.1364/ol.44.005804} {\bibfield  {journal} {\bibinfo  {journal} {Opt. Lett.}\ }\textbf {\bibinfo {volume} {44}},\ \bibinfo {pages} {5804} (\bibinfo {year} {2019}{\natexlab{a}})}\BibitemShut {NoStop}%
\bibitem [{\citenamefont {Longhi}(2019{\natexlab{b}})}]{longhi2019probing}%
  \BibitemOpen
  \bibfield  {author} {\bibinfo {author} {\bibfnamefont {S.}~\bibnamefont {Longhi}},\ }\bibfield  {title} {\bibinfo {title} {Probing non-{{Hermitian}} skin effect and non-{{Bloch}} phase transitions},\ }\href {https://doi.org/10.1103/PhysRevResearch.1.023013} {\bibfield  {journal} {\bibinfo  {journal} {Phys. Rev. Research}\ }\textbf {\bibinfo {volume} {1}},\ \bibinfo {pages} {023013} (\bibinfo {year} {2019}{\natexlab{b}})}\BibitemShut {NoStop}%
\bibitem [{\citenamefont {Xiao}\ \emph {et~al.}(2021)\citenamefont {Xiao}, \citenamefont {Deng}, \citenamefont {Wang}, \citenamefont {Wang}, \citenamefont {Yi},\ and\ \citenamefont {Xue}}]{xiao2021observation}%
  \BibitemOpen
  \bibfield  {author} {\bibinfo {author} {\bibfnamefont {L.}~\bibnamefont {Xiao}}, \bibinfo {author} {\bibfnamefont {T.}~\bibnamefont {Deng}}, \bibinfo {author} {\bibfnamefont {K.}~\bibnamefont {Wang}}, \bibinfo {author} {\bibfnamefont {Z.}~\bibnamefont {Wang}}, \bibinfo {author} {\bibfnamefont {W.}~\bibnamefont {Yi}},\ and\ \bibinfo {author} {\bibfnamefont {P.}~\bibnamefont {Xue}},\ }\bibfield  {title} {\bibinfo {title} {Observation of {{Non-Bloch Parity-Time Symmetry}} and {{Exceptional Points}}},\ }\href {https://doi.org/10.1103/PhysRevLett.126.230402} {\bibfield  {journal} {\bibinfo  {journal} {Phys. Rev. Lett.}\ }\textbf {\bibinfo {volume} {126}},\ \bibinfo {pages} {230402} (\bibinfo {year} {2021})}\BibitemShut {NoStop}%
\bibitem [{\citenamefont {Hu}\ \emph {et~al.}(2024)\citenamefont {Hu}, \citenamefont {Wang}, \citenamefont {Wang},\ and\ \citenamefont {Song}}]{hu2024geometric}%
  \BibitemOpen
  \bibfield  {author} {\bibinfo {author} {\bibfnamefont {Y.-M.}\ \bibnamefont {Hu}}, \bibinfo {author} {\bibfnamefont {H.-Y.}\ \bibnamefont {Wang}}, \bibinfo {author} {\bibfnamefont {Z.}~\bibnamefont {Wang}},\ and\ \bibinfo {author} {\bibfnamefont {F.}~\bibnamefont {Song}},\ }\bibfield  {title} {\bibinfo {title} {Geometric {{Origin}} of {{Non-Bloch P T Symmetry Breaking}}},\ }\href {https://doi.org/10.1103/PhysRevLett.132.050402} {\bibfield  {journal} {\bibinfo  {journal} {Phys. Rev. Lett.}\ }\textbf {\bibinfo {volume} {132}},\ \bibinfo {pages} {050402} (\bibinfo {year} {2024})}\BibitemShut {NoStop}%
\bibitem [{\citenamefont {McDonald}\ and\ \citenamefont {Clerk}(2020)}]{mcdonald2020exponentiallyenhanced}%
  \BibitemOpen
  \bibfield  {author} {\bibinfo {author} {\bibfnamefont {A.}~\bibnamefont {McDonald}}\ and\ \bibinfo {author} {\bibfnamefont {A.~A.}\ \bibnamefont {Clerk}},\ }\bibfield  {title} {\bibinfo {title} {Exponentially-enhanced quantum sensing with non-{{Hermitian}} lattice dynamics},\ }\href {https://doi.org/10.1038/s41467-020-19090-4} {\bibfield  {journal} {\bibinfo  {journal} {Nat Commun}\ }\textbf {\bibinfo {volume} {11}},\ \bibinfo {pages} {5382} (\bibinfo {year} {2020})}\BibitemShut {NoStop}%
\bibitem [{\citenamefont {Wanjura}\ \emph {et~al.}(2020)\citenamefont {Wanjura}, \citenamefont {Brunelli},\ and\ \citenamefont {Nunnenkamp}}]{wanjura2020topological}%
  \BibitemOpen
  \bibfield  {author} {\bibinfo {author} {\bibfnamefont {C.~C.}\ \bibnamefont {Wanjura}}, \bibinfo {author} {\bibfnamefont {M.}~\bibnamefont {Brunelli}},\ and\ \bibinfo {author} {\bibfnamefont {A.}~\bibnamefont {Nunnenkamp}},\ }\bibfield  {title} {\bibinfo {title} {Topological framework for directional amplification in driven-dissipative cavity arrays},\ }\href {https://doi.org/10.1038/s41467-020-16863-9} {\bibfield  {journal} {\bibinfo  {journal} {Nat Commun}\ }\textbf {\bibinfo {volume} {11}},\ \bibinfo {pages} {3149} (\bibinfo {year} {2020})}\BibitemShut {NoStop}%
\bibitem [{\citenamefont {Borgnia}\ \emph {et~al.}(2020)\citenamefont {Borgnia}, \citenamefont {Kruchkov},\ and\ \citenamefont {Slager}}]{borgnia2020nonhermitian}%
  \BibitemOpen
  \bibfield  {author} {\bibinfo {author} {\bibfnamefont {D.~S.}\ \bibnamefont {Borgnia}}, \bibinfo {author} {\bibfnamefont {A.~J.}\ \bibnamefont {Kruchkov}},\ and\ \bibinfo {author} {\bibfnamefont {R.-J.}\ \bibnamefont {Slager}},\ }\bibfield  {title} {\bibinfo {title} {Non-{{Hermitian Boundary Modes}} and {{Topology}}},\ }\href {https://doi.org/10.1103/PhysRevLett.124.056802} {\bibfield  {journal} {\bibinfo  {journal} {Phys. Rev. Lett.}\ }\textbf {\bibinfo {volume} {124}},\ \bibinfo {pages} {056802} (\bibinfo {year} {2020})}\BibitemShut {NoStop}%
\bibitem [{\citenamefont {Xue}\ \emph {et~al.}(2021)\citenamefont {Xue}, \citenamefont {Li}, \citenamefont {Hu}, \citenamefont {Song},\ and\ \citenamefont {Wang}}]{xue2021simple}%
  \BibitemOpen
  \bibfield  {author} {\bibinfo {author} {\bibfnamefont {W.-T.}\ \bibnamefont {Xue}}, \bibinfo {author} {\bibfnamefont {M.-R.}\ \bibnamefont {Li}}, \bibinfo {author} {\bibfnamefont {Y.-M.}\ \bibnamefont {Hu}}, \bibinfo {author} {\bibfnamefont {F.}~\bibnamefont {Song}},\ and\ \bibinfo {author} {\bibfnamefont {Z.}~\bibnamefont {Wang}},\ }\bibfield  {title} {\bibinfo {title} {Simple formulas of directional amplification from non-{{Bloch}} band theory},\ }\href {https://doi.org/10.1103/physrevb.103.l241408} {\bibfield  {journal} {\bibinfo  {journal} {Phys. Rev. B}\ }\textbf {\bibinfo {volume} {103}},\ \bibinfo {pages} {L241408} (\bibinfo {year} {2021})}\BibitemShut {NoStop}%
\bibitem [{\citenamefont {Zhu}\ \emph {et~al.}(2022)\citenamefont {Zhu}, \citenamefont {Wang}, \citenamefont {Leykam}, \citenamefont {Xue}, \citenamefont {Wang},\ and\ \citenamefont {Chong}}]{zhu2022anomalous}%
  \BibitemOpen
  \bibfield  {author} {\bibinfo {author} {\bibfnamefont {B.}~\bibnamefont {Zhu}}, \bibinfo {author} {\bibfnamefont {Q.}~\bibnamefont {Wang}}, \bibinfo {author} {\bibfnamefont {D.}~\bibnamefont {Leykam}}, \bibinfo {author} {\bibfnamefont {H.}~\bibnamefont {Xue}}, \bibinfo {author} {\bibfnamefont {Q.~J.}\ \bibnamefont {Wang}},\ and\ \bibinfo {author} {\bibfnamefont {Y.~D.}\ \bibnamefont {Chong}},\ }\bibfield  {title} {\bibinfo {title} {Anomalous {{Single-Mode Lasing Induced}} by {{Nonlinearity}} and the {{Non-Hermitian Skin Effect}}},\ }\href {https://doi.org/10.1103/PhysRevLett.129.013903} {\bibfield  {journal} {\bibinfo  {journal} {Phys. Rev. Lett.}\ }\textbf {\bibinfo {volume} {129}},\ \bibinfo {pages} {013903} (\bibinfo {year} {2022})}\BibitemShut {NoStop}%
\bibitem [{\citenamefont {Kawabata}\ \emph {et~al.}(2023)\citenamefont {Kawabata}, \citenamefont {Numasawa},\ and\ \citenamefont {Ryu}}]{kawabata2023entanglement}%
  \BibitemOpen
  \bibfield  {author} {\bibinfo {author} {\bibfnamefont {K.}~\bibnamefont {Kawabata}}, \bibinfo {author} {\bibfnamefont {T.}~\bibnamefont {Numasawa}},\ and\ \bibinfo {author} {\bibfnamefont {S.}~\bibnamefont {Ryu}},\ }\bibfield  {title} {\bibinfo {title} {Entanglement {{Phase Transition Induced}} by the {{Non-Hermitian Skin Effect}}},\ }\href {https://doi.org/10.1103/PhysRevX.13.021007} {\bibfield  {journal} {\bibinfo  {journal} {Phys. Rev. X}\ }\textbf {\bibinfo {volume} {13}},\ \bibinfo {pages} {021007} (\bibinfo {year} {2023})}\BibitemShut {NoStop}%
\bibitem [{\citenamefont {Shao}\ \emph {et~al.}(2024)\citenamefont {Shao}, \citenamefont {Geng}, \citenamefont {Liu}, \citenamefont {Lado}, \citenamefont {Chen},\ and\ \citenamefont {Xing}}]{shao2024nonHermitian}%
  \BibitemOpen
  \bibfield  {author} {\bibinfo {author} {\bibfnamefont {K.}~\bibnamefont {Shao}}, \bibinfo {author} {\bibfnamefont {H.}~\bibnamefont {Geng}}, \bibinfo {author} {\bibfnamefont {E.}~\bibnamefont {Liu}}, \bibinfo {author} {\bibfnamefont {J.~L.}\ \bibnamefont {Lado}}, \bibinfo {author} {\bibfnamefont {W.}~\bibnamefont {Chen}},\ and\ \bibinfo {author} {\bibfnamefont {D.~Y.}\ \bibnamefont {Xing}},\ }\bibfield  {title} {\bibinfo {title} {Non-{{Hermitian}} {{Moir\'e}} valley filter},\ }\href {https://doi.org/10.1103/PhysRevLett.132.156301} {\bibfield  {journal} {\bibinfo  {journal} {Phys. Rev. Lett.}\ }\textbf {\bibinfo {volume} {132}},\ \bibinfo {pages} {156301} (\bibinfo {year} {2024})}\BibitemShut {NoStop}%
\bibitem [{\citenamefont {Yokomizo}\ and\ \citenamefont {Murakami}(2019)}]{yokomizo2019nonbloch}%
  \BibitemOpen
  \bibfield  {author} {\bibinfo {author} {\bibfnamefont {K.}~\bibnamefont {Yokomizo}}\ and\ \bibinfo {author} {\bibfnamefont {S.}~\bibnamefont {Murakami}},\ }\bibfield  {title} {\bibinfo {title} {Non-{{Bloch Band Theory}} of {{Non-Hermitian Systems}}},\ }\href {https://doi.org/10.1103/PhysRevLett.123.066404} {\bibfield  {journal} {\bibinfo  {journal} {Phys. Rev. Lett.}\ }\textbf {\bibinfo {volume} {123}},\ \bibinfo {pages} {066404} (\bibinfo {year} {2019})}\BibitemShut {NoStop}%
\bibitem [{\citenamefont {Yokomizo}(2022)}]{yokomizo2022Nonbloch}%
  \BibitemOpen
  \bibfield  {author} {\bibinfo {author} {\bibfnamefont {K.}~\bibnamefont {Yokomizo}},\ }\emph {\bibinfo {title} {Non-Bloch Band Theory of Non-Hermitian Systems}},\ \href@noop {} {Ph.D. thesis},\ \bibinfo  {school} {Springer}, \bibinfo {address} {{Singapore}} (\bibinfo {year} {2022})\BibitemShut {NoStop}%
\bibitem [{\citenamefont {Yang}\ \emph {et~al.}(2020)\citenamefont {Yang}, \citenamefont {Zhang}, \citenamefont {Fang},\ and\ \citenamefont {Hu}}]{yang2020nonhermitian}%
  \BibitemOpen
  \bibfield  {author} {\bibinfo {author} {\bibfnamefont {Z.}~\bibnamefont {Yang}}, \bibinfo {author} {\bibfnamefont {K.}~\bibnamefont {Zhang}}, \bibinfo {author} {\bibfnamefont {C.}~\bibnamefont {Fang}},\ and\ \bibinfo {author} {\bibfnamefont {J.}~\bibnamefont {Hu}},\ }\bibfield  {title} {\bibinfo {title} {Non-{{Hermitian Bulk-Boundary Correspondence}} and {{Auxiliary Generalized Brillouin Zone Theory}}},\ }\href {https://doi.org/10.1103/PhysRevLett.125.226402} {\bibfield  {journal} {\bibinfo  {journal} {Phys. Rev. Lett.}\ }\textbf {\bibinfo {volume} {125}},\ \bibinfo {pages} {226402} (\bibinfo {year} {2020})}\BibitemShut {NoStop}%
\bibitem [{\citenamefont {Kawabata}\ \emph {et~al.}(2020{\natexlab{a}})\citenamefont {Kawabata}, \citenamefont {Okuma},\ and\ \citenamefont {Sato}}]{kawabata2020nonbloch}%
  \BibitemOpen
  \bibfield  {author} {\bibinfo {author} {\bibfnamefont {K.}~\bibnamefont {Kawabata}}, \bibinfo {author} {\bibfnamefont {N.}~\bibnamefont {Okuma}},\ and\ \bibinfo {author} {\bibfnamefont {M.}~\bibnamefont {Sato}},\ }\bibfield  {title} {\bibinfo {title} {Non-{{Bloch}} band theory of non-{{Hermitian Hamiltonians}} in the symplectic class},\ }\href {https://doi.org/10.1103/PhysRevB.101.195147} {\bibfield  {journal} {\bibinfo  {journal} {Phys. Rev. B}\ }\textbf {\bibinfo {volume} {101}},\ \bibinfo {pages} {195147} (\bibinfo {year} {2020}{\natexlab{a}})}\BibitemShut {NoStop}%
\bibitem [{\citenamefont {Deng}\ and\ \citenamefont {Yi}(2019)}]{deng2019nonbloch}%
  \BibitemOpen
  \bibfield  {author} {\bibinfo {author} {\bibfnamefont {T.-S.}\ \bibnamefont {Deng}}\ and\ \bibinfo {author} {\bibfnamefont {W.}~\bibnamefont {Yi}},\ }\bibfield  {title} {\bibinfo {title} {Non-{{Bloch}} topological invariants in a non-{{Hermitian}} domain wall system},\ }\href {https://doi.org/10.1103/PhysRevB.100.035102} {\bibfield  {journal} {\bibinfo  {journal} {Phys. Rev. B}\ }\textbf {\bibinfo {volume} {100}},\ \bibinfo {pages} {035102} (\bibinfo {year} {2019})}\BibitemShut {NoStop}%
\bibitem [{\citenamefont {Hu}\ \emph {et~al.}(2023{\natexlab{b}})\citenamefont {Hu}, \citenamefont {Huang}, \citenamefont {Xue},\ and\ \citenamefont {Wang}}]{hu2023non}%
  \BibitemOpen
  \bibfield  {author} {\bibinfo {author} {\bibfnamefont {Y.-M.}\ \bibnamefont {Hu}}, \bibinfo {author} {\bibfnamefont {Y.-Q.}\ \bibnamefont {Huang}}, \bibinfo {author} {\bibfnamefont {W.-T.}\ \bibnamefont {Xue}},\ and\ \bibinfo {author} {\bibfnamefont {Z.}~\bibnamefont {Wang}},\ }\bibfield  {title} {\bibinfo {title} {Non-{{Bloch}} band theory for non-{{Hermitian}} continuum systems},\ }\href@noop {} {\bibfield  {journal} {\bibinfo  {journal} {arXiv preprint arXiv:2310.08572}\ } (\bibinfo {year} {2023}{\natexlab{b}})}\BibitemShut {NoStop}%
\bibitem [{\citenamefont {Zhang}\ \emph {et~al.}(2022{\natexlab{b}})\citenamefont {Zhang}, \citenamefont {Yang},\ and\ \citenamefont {Fang}}]{zhang2022universal}%
  \BibitemOpen
  \bibfield  {author} {\bibinfo {author} {\bibfnamefont {K.}~\bibnamefont {Zhang}}, \bibinfo {author} {\bibfnamefont {Z.}~\bibnamefont {Yang}},\ and\ \bibinfo {author} {\bibfnamefont {C.}~\bibnamefont {Fang}},\ }\bibfield  {title} {\bibinfo {title} {Universal non-{{Hermitian}} skin effect in two and higher dimensions},\ }\href {https://doi.org/10.1038/s41467-022-30161-6} {\bibfield  {journal} {\bibinfo  {journal} {Nat Commun}\ }\textbf {\bibinfo {volume} {13}},\ \bibinfo {pages} {2496} (\bibinfo {year} {2022}{\natexlab{b}})}\BibitemShut {NoStop}%
\bibitem [{\citenamefont {Zhang}\ \emph {et~al.}(2023)\citenamefont {Zhang}, \citenamefont {Fang},\ and\ \citenamefont {Yang}}]{zhang2023dynamical}%
  \BibitemOpen
  \bibfield  {author} {\bibinfo {author} {\bibfnamefont {K.}~\bibnamefont {Zhang}}, \bibinfo {author} {\bibfnamefont {C.}~\bibnamefont {Fang}},\ and\ \bibinfo {author} {\bibfnamefont {Z.}~\bibnamefont {Yang}},\ }\bibfield  {title} {\bibinfo {title} {Dynamical {{Degeneracy Splitting}} and {{Directional Invisibility}} in {{Non-Hermitian Systems}}},\ }\href {https://doi.org/10.1103/PhysRevLett.131.036402} {\bibfield  {journal} {\bibinfo  {journal} {Phys. Rev. Lett.}\ }\textbf {\bibinfo {volume} {131}},\ \bibinfo {pages} {036402} (\bibinfo {year} {2023})}\BibitemShut {NoStop}%
\bibitem [{\citenamefont {Fang}\ \emph {et~al.}(2022)\citenamefont {Fang}, \citenamefont {Hu}, \citenamefont {Zhou},\ and\ \citenamefont {Ding}}]{fang2022geometrydependent}%
  \BibitemOpen
  \bibfield  {author} {\bibinfo {author} {\bibfnamefont {Z.}~\bibnamefont {Fang}}, \bibinfo {author} {\bibfnamefont {M.}~\bibnamefont {Hu}}, \bibinfo {author} {\bibfnamefont {L.}~\bibnamefont {Zhou}},\ and\ \bibinfo {author} {\bibfnamefont {K.}~\bibnamefont {Ding}},\ }\bibfield  {title} {\bibinfo {title} {Geometry-dependent skin effects in reciprocal photonic crystals},\ }\href {https://doi.org/10.1515/nanoph-2022-0211} {\bibfield  {journal} {\bibinfo  {journal} {Nanophotonics}\ }\textbf {\bibinfo {volume} {11}},\ \bibinfo {pages} {3447} (\bibinfo {year} {2022})}\BibitemShut {NoStop}%
\bibitem [{\citenamefont {Wang}\ \emph {et~al.}(2022{\natexlab{b}})\citenamefont {Wang}, \citenamefont {You},\ and\ \citenamefont {Jen}}]{wang2022nonhermitiana}%
  \BibitemOpen
  \bibfield  {author} {\bibinfo {author} {\bibfnamefont {Y.-C.}\ \bibnamefont {Wang}}, \bibinfo {author} {\bibfnamefont {J.-S.}\ \bibnamefont {You}},\ and\ \bibinfo {author} {\bibfnamefont {H.~H.}\ \bibnamefont {Jen}},\ }\bibfield  {title} {\bibinfo {title} {A non-{{Hermitian}} optical atomic mirror},\ }\href {https://doi.org/10.1038/s41467-022-32372-3} {\bibfield  {journal} {\bibinfo  {journal} {Nat Commun}\ }\textbf {\bibinfo {volume} {13}},\ \bibinfo {pages} {4598} (\bibinfo {year} {2022}{\natexlab{b}})}\BibitemShut {NoStop}%
\bibitem [{\citenamefont {Lee}\ \emph {et~al.}(2019)\citenamefont {Lee}, \citenamefont {Li},\ and\ \citenamefont {Gong}}]{lee2019hybrid}%
  \BibitemOpen
  \bibfield  {author} {\bibinfo {author} {\bibfnamefont {C.~H.}\ \bibnamefont {Lee}}, \bibinfo {author} {\bibfnamefont {L.}~\bibnamefont {Li}},\ and\ \bibinfo {author} {\bibfnamefont {J.}~\bibnamefont {Gong}},\ }\bibfield  {title} {\bibinfo {title} {Hybrid higher-order skin-topological modes in nonreciprocal systems},\ }\href {https://doi.org/10.1103/PhysRevLett.123.016805} {\bibfield  {journal} {\bibinfo  {journal} {Phys. Rev. Lett.}\ }\textbf {\bibinfo {volume} {123}},\ \bibinfo {pages} {016805} (\bibinfo {year} {2019})}\BibitemShut {NoStop}%
\bibitem [{\citenamefont {Okugawa}\ \emph {et~al.}(2020)\citenamefont {Okugawa}, \citenamefont {Takahashi},\ and\ \citenamefont {Yokomizo}}]{okugawa2020secondorder}%
  \BibitemOpen
  \bibfield  {author} {\bibinfo {author} {\bibfnamefont {R.}~\bibnamefont {Okugawa}}, \bibinfo {author} {\bibfnamefont {R.}~\bibnamefont {Takahashi}},\ and\ \bibinfo {author} {\bibfnamefont {K.}~\bibnamefont {Yokomizo}},\ }\bibfield  {title} {\bibinfo {title} {Second-order topological non-{{Hermitian}} skin effects},\ }\href {https://doi.org/10.1103/PhysRevB.102.241202} {\bibfield  {journal} {\bibinfo  {journal} {Phys. Rev. B}\ }\textbf {\bibinfo {volume} {102}},\ \bibinfo {pages} {241202} (\bibinfo {year} {2020})}\BibitemShut {NoStop}%
\bibitem [{\citenamefont {Kawabata}\ \emph {et~al.}(2020{\natexlab{b}})\citenamefont {Kawabata}, \citenamefont {Sato},\ and\ \citenamefont {Shiozaki}}]{kawabata2020higherorder}%
  \BibitemOpen
  \bibfield  {author} {\bibinfo {author} {\bibfnamefont {K.}~\bibnamefont {Kawabata}}, \bibinfo {author} {\bibfnamefont {M.}~\bibnamefont {Sato}},\ and\ \bibinfo {author} {\bibfnamefont {K.}~\bibnamefont {Shiozaki}},\ }\bibfield  {title} {\bibinfo {title} {Higher-order non-{{Hermitian}} skin effect},\ }\href {https://doi.org/10.1103/PhysRevB.102.205118} {\bibfield  {journal} {\bibinfo  {journal} {Phys. Rev. B}\ }\textbf {\bibinfo {volume} {102}},\ \bibinfo {pages} {205118} (\bibinfo {year} {2020}{\natexlab{b}})}\BibitemShut {NoStop}%
\bibitem [{\citenamefont {Zou}\ \emph {et~al.}(2021)\citenamefont {Zou}, \citenamefont {Chen}, \citenamefont {He}, \citenamefont {Bao}, \citenamefont {Lee}, \citenamefont {Sun},\ and\ \citenamefont {Zhang}}]{zou2021observation}%
  \BibitemOpen
  \bibfield  {author} {\bibinfo {author} {\bibfnamefont {D.}~\bibnamefont {Zou}}, \bibinfo {author} {\bibfnamefont {T.}~\bibnamefont {Chen}}, \bibinfo {author} {\bibfnamefont {W.}~\bibnamefont {He}}, \bibinfo {author} {\bibfnamefont {J.}~\bibnamefont {Bao}}, \bibinfo {author} {\bibfnamefont {C.~H.}\ \bibnamefont {Lee}}, \bibinfo {author} {\bibfnamefont {H.}~\bibnamefont {Sun}},\ and\ \bibinfo {author} {\bibfnamefont {X.}~\bibnamefont {Zhang}},\ }\bibfield  {title} {\bibinfo {title} {Observation of hybrid higher-order skin-topological effect in non-{{Hermitian}} topolectrical circuits},\ }\href {https://doi.org/10.1038/s41467-021-26414-5} {\bibfield  {journal} {\bibinfo  {journal} {Nat Commun}\ }\textbf {\bibinfo {volume} {12}},\ \bibinfo {pages} {7201} (\bibinfo {year} {2021})}\BibitemShut {NoStop}%
\bibitem [{\citenamefont {Zhang}\ \emph {et~al.}(2021)\citenamefont {Zhang}, \citenamefont {Tian}, \citenamefont {Jiang}, \citenamefont {Lu},\ and\ \citenamefont {Chen}}]{zhang2021observation}%
  \BibitemOpen
  \bibfield  {author} {\bibinfo {author} {\bibfnamefont {X.}~\bibnamefont {Zhang}}, \bibinfo {author} {\bibfnamefont {Y.}~\bibnamefont {Tian}}, \bibinfo {author} {\bibfnamefont {J.-H.}\ \bibnamefont {Jiang}}, \bibinfo {author} {\bibfnamefont {M.-H.}\ \bibnamefont {Lu}},\ and\ \bibinfo {author} {\bibfnamefont {Y.-F.}\ \bibnamefont {Chen}},\ }\bibfield  {title} {\bibinfo {title} {Observation of higher-order non-{{Hermitian}} skin effect},\ }\href {https://doi.org/10.1038/s41467-021-25716-y} {\bibfield  {journal} {\bibinfo  {journal} {Nat Commun}\ }\textbf {\bibinfo {volume} {12}},\ \bibinfo {pages} {5377} (\bibinfo {year} {2021})}\BibitemShut {NoStop}%
\bibitem [{\citenamefont {Schindler}\ and\ \citenamefont {Prem}(2021)}]{schindler2021dislocation}%
  \BibitemOpen
  \bibfield  {author} {\bibinfo {author} {\bibfnamefont {F.}~\bibnamefont {Schindler}}\ and\ \bibinfo {author} {\bibfnamefont {A.}~\bibnamefont {Prem}},\ }\bibfield  {title} {\bibinfo {title} {Dislocation non-{{Hermitian}} skin effect},\ }\href {https://doi.org/10.1103/PhysRevB.104.L161106} {\bibfield  {journal} {\bibinfo  {journal} {Phys. Rev. B}\ }\textbf {\bibinfo {volume} {104}},\ \bibinfo {pages} {L161106} (\bibinfo {year} {2021})}\BibitemShut {NoStop}%
\bibitem [{\citenamefont {Sun}\ \emph {et~al.}(2021)\citenamefont {Sun}, \citenamefont {Zhu},\ and\ \citenamefont {Hughes}}]{sun2021geometric}%
  \BibitemOpen
  \bibfield  {author} {\bibinfo {author} {\bibfnamefont {X.-Q.}\ \bibnamefont {Sun}}, \bibinfo {author} {\bibfnamefont {P.}~\bibnamefont {Zhu}},\ and\ \bibinfo {author} {\bibfnamefont {T.~L.}\ \bibnamefont {Hughes}},\ }\bibfield  {title} {\bibinfo {title} {Geometric {{Response}} and {{Disclination-Induced Skin Effects}} in {{Non-Hermitian Systems}}},\ }\href {https://doi.org/10.1103/PhysRevLett.127.066401} {\bibfield  {journal} {\bibinfo  {journal} {Phys. Rev. Lett.}\ }\textbf {\bibinfo {volume} {127}},\ \bibinfo {pages} {066401} (\bibinfo {year} {2021})}\BibitemShut {NoStop}%
\bibitem [{\citenamefont {Kawabata}\ \emph {et~al.}(2021)\citenamefont {Kawabata}, \citenamefont {Shiozaki},\ and\ \citenamefont {Ryu}}]{kawabata2021topological}%
  \BibitemOpen
  \bibfield  {author} {\bibinfo {author} {\bibfnamefont {K.}~\bibnamefont {Kawabata}}, \bibinfo {author} {\bibfnamefont {K.}~\bibnamefont {Shiozaki}},\ and\ \bibinfo {author} {\bibfnamefont {S.}~\bibnamefont {Ryu}},\ }\bibfield  {title} {\bibinfo {title} {Topological {{Field Theory}} of {{Non-Hermitian Systems}}},\ }\href {https://doi.org/10.1103/PhysRevLett.126.216405} {\bibfield  {journal} {\bibinfo  {journal} {Phys. Rev. Lett.}\ }\textbf {\bibinfo {volume} {126}},\ \bibinfo {pages} {216405} (\bibinfo {year} {2021})}\BibitemShut {NoStop}%
\bibitem [{\citenamefont {Wang}\ \emph {et~al.}(2024)\citenamefont {Wang}, \citenamefont {Song},\ and\ \citenamefont {Wang}}]{wang2024amoeba}%
  \BibitemOpen
  \bibfield  {author} {\bibinfo {author} {\bibfnamefont {H.-Y.}\ \bibnamefont {Wang}}, \bibinfo {author} {\bibfnamefont {F.}~\bibnamefont {Song}},\ and\ \bibinfo {author} {\bibfnamefont {Z.}~\bibnamefont {Wang}},\ }\bibfield  {title} {\bibinfo {title} {Amoeba formulation of non-{{Bloch}} band theory in arbitrary dimensions},\ }\href {https://doi.org/10.1103/PhysRevX.14.021011} {\bibfield  {journal} {\bibinfo  {journal} {Phys. Rev. X}\ }\textbf {\bibinfo {volume} {14}},\ \bibinfo {pages} {021011} (\bibinfo {year} {2024})}\BibitemShut {NoStop}%
\bibitem [{\citenamefont {Hu}(2024)}]{hu2024topological}%
  \BibitemOpen
  \bibfield  {author} {\bibinfo {author} {\bibfnamefont {H.}~\bibnamefont {Hu}},\ }\bibfield  {title} {\bibinfo {title} {Topological origin of non-{{Hermitian}} skin effect in higher dimensions and uniform spectra},\ }\bibfield  {journal} {\bibinfo  {journal} {Science Bulletin}\ }\href {https://doi.org/https://doi.org/10.1016/j.scib.2024.07.022} {https://doi.org/10.1016/j.scib.2024.07.022} (\bibinfo {year} {2024})\BibitemShut {NoStop}%
\bibitem [{\citenamefont {Jiang}\ and\ \citenamefont {Lee}(2023)}]{jiang2023dimensional}%
  \BibitemOpen
  \bibfield  {author} {\bibinfo {author} {\bibfnamefont {H.}~\bibnamefont {Jiang}}\ and\ \bibinfo {author} {\bibfnamefont {C.~H.}\ \bibnamefont {Lee}},\ }\bibfield  {title} {\bibinfo {title} {Dimensional {{Transmutation}} from {{Non-Hermiticity}}},\ }\href {https://doi.org/10.1103/PhysRevLett.131.076401} {\bibfield  {journal} {\bibinfo  {journal} {Phys. Rev. Lett.}\ }\textbf {\bibinfo {volume} {131}},\ \bibinfo {pages} {076401} (\bibinfo {year} {2023})}\BibitemShut {NoStop}%
\bibitem [{\citenamefont {Yokomizo}\ and\ \citenamefont {Murakami}(2023)}]{yokomizo2023nonbloch}%
  \BibitemOpen
  \bibfield  {author} {\bibinfo {author} {\bibfnamefont {K.}~\bibnamefont {Yokomizo}}\ and\ \bibinfo {author} {\bibfnamefont {S.}~\bibnamefont {Murakami}},\ }\bibfield  {title} {\bibinfo {title} {Non-{{Bloch}} bands in two-dimensional non-{{Hermitian}} systems},\ }\href {https://doi.org/10.1103/PhysRevB.107.195112} {\bibfield  {journal} {\bibinfo  {journal} {Phys. Rev. B}\ }\textbf {\bibinfo {volume} {107}},\ \bibinfo {pages} {195112} (\bibinfo {year} {2023})}\BibitemShut {NoStop}%
\bibitem [{\citenamefont {Zhang}\ \emph {et~al.}(2024)\citenamefont {Zhang}, \citenamefont {Yang},\ and\ \citenamefont {Sun}}]{kai2024edge}%
  \BibitemOpen
  \bibfield  {author} {\bibinfo {author} {\bibfnamefont {K.}~\bibnamefont {Zhang}}, \bibinfo {author} {\bibfnamefont {Z.}~\bibnamefont {Yang}},\ and\ \bibinfo {author} {\bibfnamefont {K.}~\bibnamefont {Sun}},\ }\bibfield  {title} {\bibinfo {title} {Edge theory of non-{{Hermitian}} skin modes in higher dimensions},\ }\href {https://doi.org/10.1103/PhysRevB.109.165127} {\bibfield  {journal} {\bibinfo  {journal} {Phys. Rev. B}\ }\textbf {\bibinfo {volume} {109}},\ \bibinfo {pages} {165127} (\bibinfo {year} {2024})}\BibitemShut {NoStop}%
\bibitem [{\citenamefont {Xiong}\ \emph {et~al.}()\citenamefont {Xiong}, \citenamefont {Xing},\ and\ \citenamefont {Hu}}]{xiong2024non}%
  \BibitemOpen
  \bibfield  {author} {\bibinfo {author} {\bibfnamefont {Y.}~\bibnamefont {Xiong}}, \bibinfo {author} {\bibfnamefont {Z.-Y.}\ \bibnamefont {Xing}},\ and\ \bibinfo {author} {\bibfnamefont {H.}~\bibnamefont {Hu}},\ }\href@noop {} {\bibinfo {title} {Non-{{Hermitian}} skin effect in arbitrary dimensions: non-{{Bloch}} band theory and classification}},\ \Eprint {https://arxiv.org/abs/2407.01296} {arXiv:2407.01296} \BibitemShut {NoStop}%
\bibitem [{\citenamefont {Zhang}\ \emph {et~al.}()\citenamefont {Zhang}, \citenamefont {Shu},\ and\ \citenamefont {Sun}}]{zhang2024algebraic}%
  \BibitemOpen
  \bibfield  {author} {\bibinfo {author} {\bibfnamefont {K.}~\bibnamefont {Zhang}}, \bibinfo {author} {\bibfnamefont {C.}~\bibnamefont {Shu}},\ and\ \bibinfo {author} {\bibfnamefont {K.}~\bibnamefont {Sun}},\ }\href@noop {} {\bibinfo {title} {Algebraic non-{{Hermitian}} skin effect and unified non-bloch band theory in arbitrary dimensions}},\ \Eprint {https://arxiv.org/abs/2406.06682} {arXiv:2406.06682} \BibitemShut {NoStop}%
\bibitem [{\citenamefont {Xu}\ \emph {et~al.}()\citenamefont {Xu}, \citenamefont {Pang}, \citenamefont {Zhang},\ and\ \citenamefont {Yang}}]{xu2023two}%
  \BibitemOpen
  \bibfield  {author} {\bibinfo {author} {\bibfnamefont {Z.}~\bibnamefont {Xu}}, \bibinfo {author} {\bibfnamefont {B.}~\bibnamefont {Pang}}, \bibinfo {author} {\bibfnamefont {K.}~\bibnamefont {Zhang}},\ and\ \bibinfo {author} {\bibfnamefont {Z.}~\bibnamefont {Yang}},\ }\href@noop {} {\bibinfo {title} {Two-dimensional asymptotic generalized {{Brillouin}} zone conjecture}},\ \Eprint {https://arxiv.org/abs/2311.16868} {arXiv:2311.16868} \BibitemShut {NoStop}%
\bibitem [{Note1()}]{Note1}%
  \BibitemOpen
  \bibinfo {note} {Without extra explanations, the $\omega $ in $\omega -H$ denotes the product between $\omega $ and the identity matrix.}\BibitemShut {Stop}%
\bibitem [{Note2()}]{Note2}%
  \BibitemOpen
  \bibinfo {note} {According to Ref. \cite {kunst2018biorthogonal}, the bi-orthogonality between right eigenstates $|nR\rangle $ and left eigenstates $|nL\rangle $ of a non-Hermitian Hamiltonian $H$ makes sure that $\langle x|nR\rangle \langle nL|x\rangle $ is extended for the bulk states. One can use this property to show that the diagonal elements of the Green's function $G(\omega )=(\omega -H)^{-1}$ are at order $\protect \mathcal {O}(1)$.}\BibitemShut {Stop}%
\bibitem [{\citenamefont {Li}\ \emph {et~al.}(2020)\citenamefont {Li}, \citenamefont {Lee}, \citenamefont {Mu},\ and\ \citenamefont {Gong}}]{li2020critical}%
  \BibitemOpen
  \bibfield  {author} {\bibinfo {author} {\bibfnamefont {L.}~\bibnamefont {Li}}, \bibinfo {author} {\bibfnamefont {C.~H.}\ \bibnamefont {Lee}}, \bibinfo {author} {\bibfnamefont {S.}~\bibnamefont {Mu}},\ and\ \bibinfo {author} {\bibfnamefont {J.}~\bibnamefont {Gong}},\ }\bibfield  {title} {\bibinfo {title} {Critical non-{{Hermitian}} skin effect},\ }\href {https://doi.org/10.1038/s41467-020-18917-4} {\bibfield  {journal} {\bibinfo  {journal} {Nat Commun}\ }\textbf {\bibinfo {volume} {11}},\ \bibinfo {pages} {5491} (\bibinfo {year} {2020})}\BibitemShut {NoStop}%
\bibitem [{SM()}]{SM}%
  \BibitemOpen
  \href@noop {} {}\bibinfo {howpublished} {See Supplemental Material for details.}\BibitemShut {Stop}%
\bibitem [{\citenamefont {Wang}\ \emph {et~al.}(pear)\citenamefont {Wang}, \citenamefont {Song},\ and\ \citenamefont {Wang}}]{wang2024singular}%
  \BibitemOpen
  \bibfield  {author} {\bibinfo {author} {\bibfnamefont {H.-Y.}\ \bibnamefont {Wang}}, \bibinfo {author} {\bibfnamefont {F.}~\bibnamefont {Song}},\ and\ \bibinfo {author} {\bibfnamefont {Z.}~\bibnamefont {Wang}},\ }\href@noop {} {} (\bibinfo {year} {to appear})\BibitemShut {NoStop}%
\bibitem [{\citenamefont {Lin}\ \emph {et~al.}(2023)\citenamefont {Lin}, \citenamefont {Yi},\ and\ \citenamefont {Xue}}]{lin2023manipulating}%
  \BibitemOpen
  \bibfield  {author} {\bibinfo {author} {\bibfnamefont {Q.}~\bibnamefont {Lin}}, \bibinfo {author} {\bibfnamefont {W.}~\bibnamefont {Yi}},\ and\ \bibinfo {author} {\bibfnamefont {P.}~\bibnamefont {Xue}},\ }\bibfield  {title} {\bibinfo {title} {Manipulating directional flow in a two-dimensional photonic quantum walk under a synthetic magnetic field},\ }\href {https://doi.org/10.1038/s41467-023-42045-4} {\bibfield  {journal} {\bibinfo  {journal} {Nat Commun}\ }\textbf {\bibinfo {volume} {14}},\ \bibinfo {pages} {6283} (\bibinfo {year} {2023})}\BibitemShut {NoStop}%
\bibitem [{\citenamefont {Xue}\ \emph {et~al.}(2024)\citenamefont {Xue}, \citenamefont {Lin}, \citenamefont {Wang}, \citenamefont {Xiao}, \citenamefont {Longhi},\ and\ \citenamefont {Yi}}]{xue2024self}%
  \BibitemOpen
  \bibfield  {author} {\bibinfo {author} {\bibfnamefont {P.}~\bibnamefont {Xue}}, \bibinfo {author} {\bibfnamefont {Q.}~\bibnamefont {Lin}}, \bibinfo {author} {\bibfnamefont {K.}~\bibnamefont {Wang}}, \bibinfo {author} {\bibfnamefont {L.}~\bibnamefont {Xiao}}, \bibinfo {author} {\bibfnamefont {S.}~\bibnamefont {Longhi}},\ and\ \bibinfo {author} {\bibfnamefont {W.}~\bibnamefont {Yi}},\ }\bibfield  {title} {\bibinfo {title} {Self acceleration from spectral geometry in dissipative quantum-walk dynamics},\ }\href {https://doi.org/10.1038/s41467-024-48815-y} {\bibfield  {journal} {\bibinfo  {journal} {Nat Commun}\ }\textbf {\bibinfo {volume} {15}},\ \bibinfo {pages} {4381} (\bibinfo {year} {2024})}\BibitemShut {NoStop}%
\bibitem [{\citenamefont {Shu}\ \emph {et~al.}()\citenamefont {Shu}, \citenamefont {Zhang},\ and\ \citenamefont {Sun}}]{shu2024ultra}%
  \BibitemOpen
  \bibfield  {author} {\bibinfo {author} {\bibfnamefont {C.}~\bibnamefont {Shu}}, \bibinfo {author} {\bibfnamefont {K.}~\bibnamefont {Zhang}},\ and\ \bibinfo {author} {\bibfnamefont {K.}~\bibnamefont {Sun}},\ }\href@noop {} {\bibinfo {title} {Ultra spectral sensitivity and non-local bi-impurity bound states from quasi-long-range non-{{Hermitian}} skin modes}},\ \Eprint {https://arxiv.org/abs/2409.13623} {arXiv:2409.13623} \BibitemShut {NoStop}%
\end{thebibliography}%

\end{document}